\documentclass[journal=jcp, preprint, manuscript=article, email=true]{achemso}
\usepackage{graphicx} 
\usepackage{import}
\usepackage{xcolor}
\usepackage{amsmath} 
\usepackage{amssymb}
\usepackage{booktabs}
\usepackage{tabularx}
\usepackage[colorlinks]{hyperref}
\usepackage{tikz}
\usepackage{subcaption}
\hypersetup{
    colorlinks=true,
    linkcolor=black,    
    citecolor=blue,   
    urlcolor=blue,  
}
\newcommand{\namedsection}[1]{\section{#1}\addcontentsline{toc}{section}{#1}}
\newcommand{\namedsubsection}[1]{\subsection{#1}\addcontentsline{toc}{subsection}{#1}}  

\title{Citrine Informatics: Chemical \& Materials Development Platform}
\author{Maxwell C. Venetos}
\author{Steven J. Brown}
\author{Kenneth Kroenlein}
\author{Steven K. Kauwe}
\author{James E. Saal}
\author{Marco Musto}
\author{Matthew D. Gerboth}
\author{Kyle D. Miller}
\author{Gregory J. Mulholland}
\email{gmulholland@citrine.io}
\affiliation{Citrine Informatics, Redwood City, California 94063, United States}

\begin{document}

\maketitle
\let\thefootnote\relax\footnotetext{Copyright \textcopyright  ~2026 Citrine Informatics, Inc. All rights reserved.}
\begin{abstract}
Today the Citrine Platform regularly powers data-driven materials discovery across industries, having moved beyond one-off demonstrations into routine industrial practice. Getting there required solving a core set of recurring obstacles: experimental data are scarce, costly, and published in formats that resist reuse; conventional accuracy metrics overstate model performance under the extrapolative conditions that define discovery; and realistic design spaces are bounded by physics, manufacturability, supply, and cost. Developed over more than a decade as an integrated response to these obstacles, the Citrine Platform is organized as four cooperating stages within a closed sequential learning loop. Stage~1 ingests and featurizes data through the Graphical Expression of Materials Data (GEMD) model, which treats process history, measurement uncertainty, and provenance as first-class features. Stage~2 builds machine learning models with well-calibrated uncertainty, including multivariate prediction intervals for correlated objectives, and validates them with extrapolative cross-validation and dynamic discovery metrics rather than random held-out splits. Stage~3 encodes compositional, physical, processing, and economic constraints directly into the design space, and Stage~4 applies the FUELS sequential learning framework with uncertainty-aware acquisition functions to navigate large constrained spaces under tight evaluation budgets. Published case studies spanning organic semiconductors, autonomous nanoparticle synthesis, and benchmark optimization tasks demonstrate two- to nine-fold reductions in experimental effort relative to random search, illustrating a stack in which data, modeling, and design-space layers continuously co-evolve.
\end{abstract}


{\footnotesize\tableofcontents}

\namedsection{Introduction}
\begin{figure}
    \centering
    \includegraphics[width=0.75\linewidth]{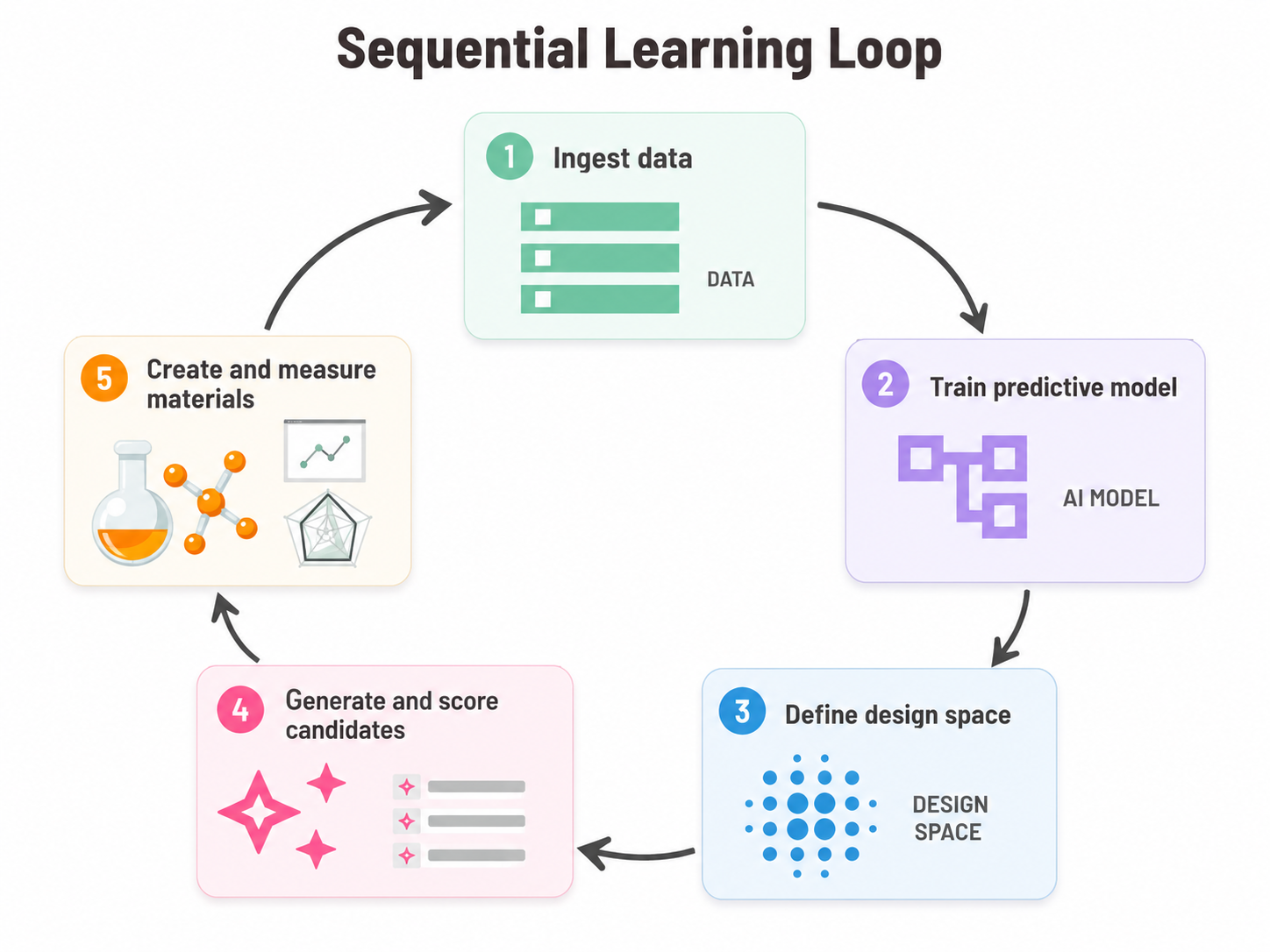}
    \caption{The sequential learning loop underlying the Citrine Platform. Stages~1–4: data ingestion, predictive model training, design-space definition, and candidate generation and scoring are supported by the platform and correspond to the four stage sections of this paper. Stage~5, in which recommended candidates are synthesized and characterized, is carried out by the user; the resulting measurements are returned to Stage~1, where they augment the training set and refine subsequent iterations. Sustained model improvement therefore depends on the user closing the loop through Stage~5, making experimental follow-through an integral part of the workflow rather than a downstream activity.}
    \label{fig:SL_fig}
\end{figure}
The Materials Genome Initiative and the broader push toward data-driven materials discovery have produced a growing number of computational databases, autonomous experimental hardware, and machine learning (ML) methods aimed at compressing the historic decades-long cycle from materials invention to deployment. High-throughput density functional theory (DFT) repositories such as the Materials Project~\cite{jain2013mp}, the Open Quantum Materials Database~\cite{saal2013oqmd,shen2022oqmd}, and AFLOW~\cite{curtarolo2012aflow} now expose hundreds of thousands of computed structures and properties to the community. Surveys of materials informatics methods document a similar acceleration on the algorithmic side~\cite{ramprasad2017,jain2024ml}, and reviews of experimentally or computationally validated ML-driven discoveries~\cite{saal2020} now span material classes ranging from thermoelectrics and alloys to organic semiconductors. Robotic experimentation and autonomous laboratories are beginning to close the loop between prediction and synthesis~\cite{tabor2018automation,szymanski2023autonomous,raccuglia2016}. Despite this momentum, translating individual demonstrations into reliable end-to-end discovery workflows in industrial research and development remains difficult.

Three obstacles recur whenever materials informatics is moved from a successful publication into a sustained discovery program. The first is data: experimental materials data are expensive to generate, with the cost of characterizing a single novel material running into tens of thousands of US dollars~\cite{lookman2019al,tabor2018automation}, and they are typically published in formats that resist programmatic reuse. The FAIR principles~\cite{wilkinson2016fair,scheffler2022fair,brinson2024fair} articulate the data-infrastructure requirements that any cross-study aggregation must satisfy, but achieving findability, accessibility, interoperability, and reusability across heterogeneous experimental and computational sources remains an open problem. Even the major DFT databases disagree with one another at levels comparable to DFT-versus-experiment errors when their records are aligned and compared~\cite{hegde2023prm}, and standardizing multimodal experimental data across institutions is an active area of research~\cite{allec2024immi}. The second obstacle is statistical: conventional ML metrics such as RMSE and $R^2$ overestimate model performance under the extrapolative conditions that define discovery~\cite{meredig2018_rsc}, and they correlate poorly with a model's actual ability to identify high-performing candidates in a sequential learning campaign~\cite{borg2023}. Well-calibrated uncertainty estimates, including the multivariate intervals required when correlated objectives must be optimized jointly~\cite{folie2023mlst}, are thus a prerequisite rather than an embellishment. The third obstacle is that realistic design spaces are not unconstrained: they are bounded by physics, manufacturability, supply, and cost. Screen-and-rank pipelines that ignore such constraints rarely produce candidates that survive engineering review.

The Citrine Platform has been developed over more than a decade as a coordinated response to these obstacles. At the data layer, the platform combines a hosted infrastructure for ingestion, storage, and search~\cite{omara2016,hill2016mrs,hill2018chapter} with a sequence of progressively richer schemas, from the Physical Information File~\cite{citrinePIF2016} to the Graphical Expression of Materials Data (GEMD) model~\cite{citrine2020gemd}. This layer makes process history, measurement uncertainty, and provenance first-class features of the data model rather than optional metadata. At the modeling layer, the platform implements the FUELS sequential learning framework~\cite{Ling2017} together with the open-source \texttt{Lolo} library~\cite{lolo} to provide well-calibrated random-forest uncertainty estimates, extending to multivariate prediction intervals for correlated objectives~\cite{folie2023mlst}; it adopts evaluation protocols such as Leave-One-Cluster-Out cross-validation~\cite{meredig2018_rsc} and dynamic discovery-based metrics~\cite{borg2023} that measure model utility under the conditions in which materials are actually discovered. The resulting platform has been applied across a wide range of industrial materials problems~\cite{meredig2017indus}.

This paper organizes the Citrine Platform as a sequence of four stages within the broader sequential learning loop illustrated in Fig.~\ref{fig:SL_fig}. The platform supports stages 1-4; the fifth step of the loop, synthesis and characterization of the recommended candidates, is carried out by the user, who returns the resulting measurements to Stage~1 and thereby sustains the iterative improvement on which sequential learning depends. Stage~1 covers data ingestion and featurization, including the GEMD data model and the platform's featurization pipeline, where FAIR considerations and data-quality concerns are made operational. Stage~2 covers ML model construction, including automated algorithm selection, well-calibrated uncertainty quantification, and validation protocols suited to extrapolative discovery. Stage~3 covers design space definition: the encoding of domain-knowledge constraints (compositional, physical, processing, and economic) and the generation of candidate materials that satisfy them. Stage~4 covers optimization within the design space, presenting the sequential learning framework, the acquisition functions used to balance exploration against exploitation, and the engineering choices that make the framework efficient and scalable in production use.

The remainder of the paper is organized accordingly. The four stage sections are presented in order, each motivating the design choices made within the platform and relating them to the broader materials informatics literature. The Applications section then summarizes published case studies in which FUELS-based sequential learning has been used to discover new materials and to optimize manufacturing processes, and the Conclusions section synthesizes the lessons drawn from this body of work and identifies open challenges for the next generation of materials informatics platforms.

\namedsection{Stage 1: Data Ingestion and Featurization}
\namedsubsection{Data Challenge in Materials Science}
\subsubsection{Data Quality in Machine Learning Contexts}

Traditional machine learning (ML) endeavors rely on a canonical set of metrics to evaluate the utility of a dataset: size, diversity, balance, and consistency\cite{mohammed2025dataquality,zhou2024dqsurvey}. A large corpus is generally considered a prerequisite for robust model training, but volume alone is insufficient. Diversity ensures that examples cover the full breadth of the problem space, reducing the risk of overfitting to a narrow region\cite{gong2019diversity}. Balance prevents systematic bias toward the most frequently observed classes or conditions, a concern particularly acute in classification tasks\cite{he2009imbalanced}. Consistency in measurement methodology, feature definition, and data collection protocols is necessary to prevent confounding variables from misleading the model\cite{sambasivan2021cascades,geirhos2020shortcut}.

In materials science, each of these concerns is acutely present. Collecting large training sets is often prohibitively expensive; characterizing a single novel material can cost tens of thousands of US dollars~\cite{lookman2019al,tabor2018automation,raccuglia2016}. Individual studies are typically designed with a specific optimization objective in mind, so the resulting data tend to cluster narrowly in composition or processing space rather than spanning it broadly. Poorly performing materials are rarely investigated, producing fundamental imbalance in any retrospectively assembled corpus. Aggregating results from multiple studies can improve diversity, but doing so introduces inconsistency in representation and measurement uncertainty, and often fails to overcome the inherent clustering of single-material optimization campaigns~\cite{ramprasad2017,hill2016mrs}. Large-scale density functional theory (DFT) databases such as the Materials Project~\cite{jain2013mp} and the Open Quantum Materials Database (OQMD)~\cite{saal2013oqmd,shen2022oqmd} have been developed to provide large, internally consistent, and compositionally diverse datasets; however, these sources describe computed ground-state structures rather than experimentally realized materials, limiting their generalizability. All of these considerations are predicated on the more fundamental concern of locating and conditioning relevant data.

\subsubsection{The FAIR Framework in Materials Science}

A researcher building a materials ML model faces two distinct questions about any candidate dataset. The first, addressed by the metrics above, is whether the dataset would be statistically useful if incorporated: is it large enough, diverse enough, and consistent enough to improve the model? The second question is prerequisite to the first: can the data be found, accessed, and understood outside its original context, and can it be combined with other sources without loss of meaning? A researcher generating data faces the symmetric challenge: how should a dataset be described and structured so that future practitioners can determine whether it is relevant to their needs and use it with confidence? The FAIR framework, introduced by Wilkinson and colleagues~\cite{wilkinson2016fair}, was developed to articulate precisely this second class of questions. FAIR (Findable, Accessible, Interoperable, and Reusable) provides a vocabulary for diagnosing whether data infrastructure supports discovery and reuse, and has gained significant traction in the materials informatics community as a practical standard for data sharing~\cite{draxl2019nomad,scheffler2022fair,brinson2024fair}.

\begin{figure}
    \centering
    \includegraphics[width=\linewidth]{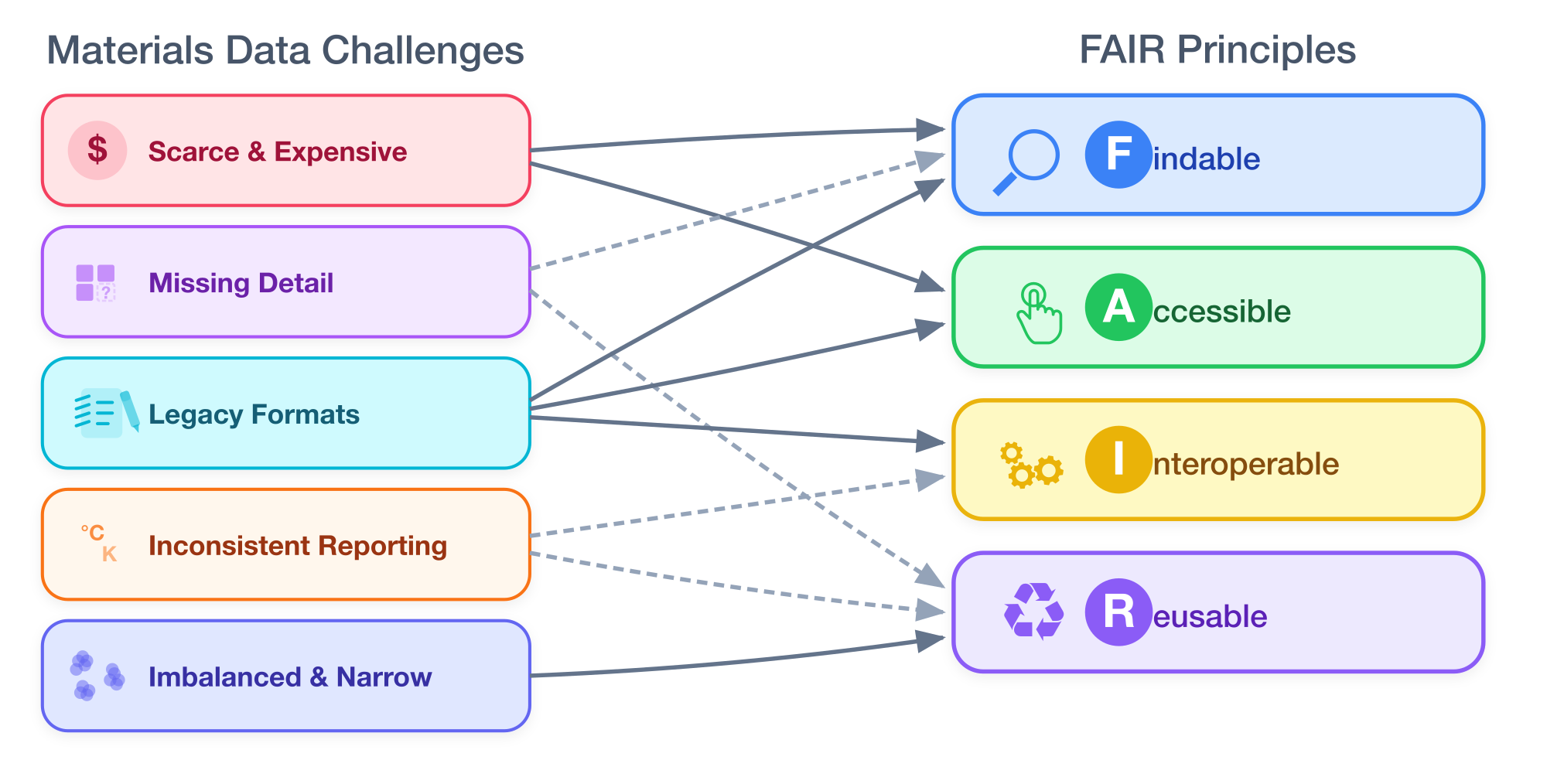}
    \caption{Exemplar relationships between materials science data challenges and the FAIR data principles. For example: the scarcity and expense of materials data results in difficulties identifying data that is relevant to a given need and many data holders will be unwilling to share their data. FAIR icons adapted from~\cite{sangyapundir2016fair}.}
    \label{fig:fair}
\end{figure}

Although the four FAIR dimensions describe what any careful researcher would naturally expect from well-curated data, they were articulated primarily to address the challenge of machine-actionability: the ability of computational systems to discover, access, integrate, and analyze data with minimal human intervention. An experienced researcher navigating an unfamiliar dataset can draw on context, domain knowledge, and intuition to resolve ambiguities, inferring from surrounding context that an unlabeled column is probably hardness in GPa, or recognizing that two differently named measurements are functionally equivalent. A computational pipeline cannot. FAIR data infrastructure is, in this sense, the discipline of making implicit scientific knowledge explicit enough for a computer to act on it reliably. In doing so, it also makes that knowledge more transparent and reproducible for human researchers.

\textbf{Findable} data are well-indexed and discoverable, both within a local collection and within the broader scientific ecosystem. Locally, findability corresponds to the ability to retrieve experimental records matching specified criteria from a LIMS or materials database. Globally, it depends on the assignment of persistent identifiers that can be cited and resolved programmatically across repositories~\cite{blaiszik2016mdf}. Examples of persistent identifiers include DOIs for datasets, InChI keys for chemical structures, and catalog numbers for commercial reagents. The persistent challenge in materials science lies in linking individual data points within publications to the underlying raw data from which they were derived.

\textbf{Accessible} data has clearly documented processes for acquisition by  any authorized user---human or machine. For digital records, accessibility encompasses authentication requirements and technical consumption barriers: supported file formats, API documentation, and software dependencies. For archival materials such as laboratory notebooks, accessibility additionally involves the substantial effort of transforming unstructured, handwritten information into machine-readable form. A common accessibility failure in materials science is the publication of data in tables or figures within journal articles, which are technically readable by humans but practically inaccessible to automated pipelines without significant manual transcription or optical character recognition.

\textbf{Interoperable} data have internal structures and semantic meanings which are mutually consistent, enabling joint analysis across sources. Incompatibilities arise at several levels of severity. Superficial syntactic differences, such as one source using "FCC" and another "face-centered cubic," or comma- versus semicolon-delimited CSV conventions, introduce friction but are generally resolvable with preprocessing. Differences in units of measure may be straightforward to reconcile (interconverting temperatures among Kelvin, Celsius, and Fahrenheit) or may require auxiliary physical information (relating dynamic and kinematic viscosity requires knowledge of the fluid's density). Some quantities are fundamentally incompatible even when they covary strongly, such as Motor Octane Number and Road Octane Number\cite{mittal2008octane}. At the most severe level, different experimental and computational paradigms require categorically different input specifications: the parameters governing a DFT calculation differ fundamentally from those of an X-ray diffraction experiment, even though both yield lattice constants as outputs. Reconciling two disparate datasets is always challenging; reconciling three or more is often impossible without information loss or unverifiable modeling assumptions.

\textbf{Reusable} data provide sufficient statistical constraint to be useful in contexts beyond that of their original purpose. Reusability depends critically on the quality and completeness of metadata: information not directly targeted for modeling but necessary for a domain expert to assess a datum's applicability. Instrument calibration records, reagent purity specifications, simulation convergence criteria, and environmental conditions during measurement are all examples. An expert integrating a new data point holistically evaluates such metadata when assigning an effective uncertainty and deciding how much statistical weight the point should carry. Without well-structured metadata, data consumers cannot reliably distinguish high-leverage observations from outliers, undermining principled model construction.

Taken together, the four FAIR dimensions set a demanding standard that few existing materials datasets meet comprehensively. The interoperability and reusability dimensions are particularly challenging, as they require not only consistent formatting but also shared semantic structures: controlled vocabularies, explicit uncertainty representations, and documented provenance. The GEMD data model, described in the following subsection, was designed with precisely these requirements in mind.

\namedsubsection{GEMD Data Model}
Citrine developed and maintains the open-source Graphical Expression of Materials Data (GEMD) model~\cite{citrine2020gemd} to represent the complete provenance of a material as a structured, interoperable graph. GEMD's design represents a significant advance over its predecessor, the Physical Information File (PIF)~\cite{citrinePIF2016}, and over ontology-based approaches such as the European Materials Modelling Ontology (EMMO)~\cite{goldbeck2019emmo,emmc2024emmo} and the PMD Core Ontology~\cite{bayerlein2024pmd}, combining the expressiveness of a rich semantic model with the compatibility of a schema-flexible, JSON-native format.

\subsubsection{Core Object Types and the Materials Graph}

GEMD organizes materials data around four core object types: \textbf{Materials}, \textbf{Processes}, \textbf{Measurements}, and \textbf{Ingredients}. These categories map directly onto the physical activities of a materials laboratory. A Process consumes zero or more input Materials (mediated through Ingredient objects that annotate quantity and role) and produces exactly one output Material. A Measurement characterizes a Material by recording properties along with the parameters and conditions under which they were obtained. Together, these objects represent the chronology of how a sample came to be. The linked objects form a directed acyclic graph (DAG) in which the terminal Material is the primary artifact of interest, and its complete history can be reconstructed by traversing the graph backward through successive processing steps to the starting precursor materials. GEMD thereby links records describing processing, structure, properties, and performance in a single traversable object.

\begin{figure}
    \centering
    \includegraphics[width=\linewidth]{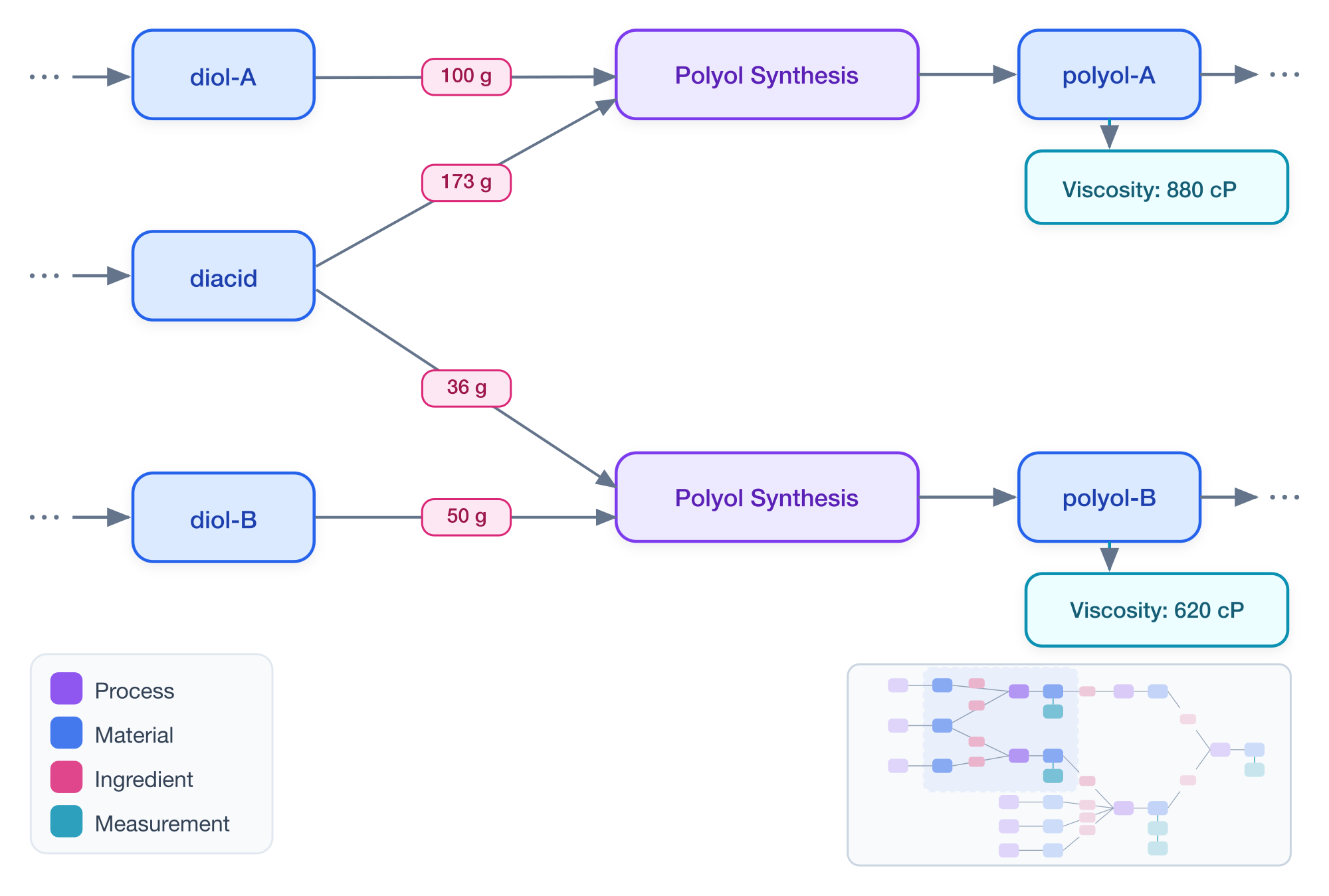}
    \caption{Schematic representation of a GEMD material history. The main panel shows an example graph in which precursor materials are combined through processing steps to produce output materials, which are then characterized by measurements. The inset illustrates how a single material history connects to a larger network of upstream and downstream processing stages.}
    \label{fig:gemd-schematic}
\end{figure}

This graph structure offers a decisive advantage over the tabular representations prevalent in materials databases~\cite{ramprasad2017,hill2016mrs}. Tabular formats collapse the process history into a flat row of features, losing the causal relationships between processing steps. The GEMD graph preserves these relationships explicitly, enabling queries such as ``identify all materials that underwent a sintering step above 1500~K as a second processing stage,'' a query that is impossible to answer from a conventional spreadsheet. It also supports multi-step synthesis workflows natively: the output Material of one Process becomes an Ingredient in a subsequent Process, forming a chain of arbitrary depth.

\subsubsection{The Template--Spec--Run Hierarchy}

Perhaps GEMD's most conceptually distinctive feature is its three-level hierarchy for every object type: \textbf{Templates}, \textbf{Specs}, and \textbf{Runs}.

A \textbf{Template} defines the schema of a class of objects, specifying which attributes are expected and bounding their valid ranges. A \textbf{Spec} represents the intent to perform a specific experiment or process: the planned temperature, the target composition, the intended measurement protocol. A \textbf{Run} records what actually happened: the measured temperature, the as-synthesized composition, the observed property values.

\begin{figure}
    \centering
    \includegraphics[width=\linewidth]{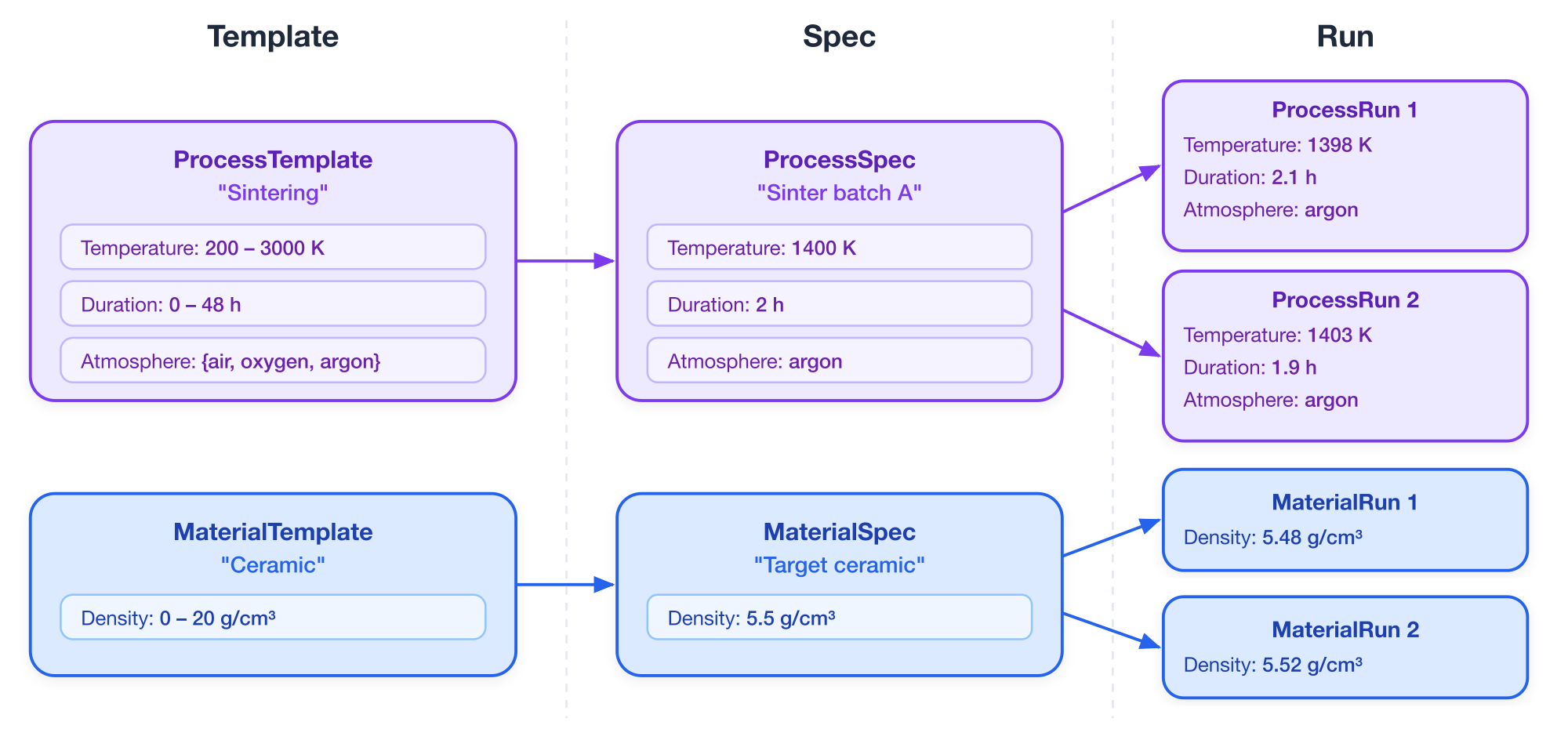}
    \caption{The Template--Spec--Run hierarchy illustrated for a sintering process and its output ceramic material. Templates define the schema and valid attribute ranges, Specs express the intended experimental conditions, and Runs record the values actually realized. A single Spec may be associated with multiple Runs, making the deviation between intent and outcome a queryable quantity.}
    \label{fig:tsr-hierarchy}
\end{figure}

This distinction resolves an ambiguity endemic to materials databases: are the recorded values what was planned or what was observed? GEMD answers this question structurally. A single Process Spec (e.g., ``sinter at 1400\textdegree C for two hours'') can be associated with multiple Process Runs, each recording the actual conditions realized during a particular experimental execution. The deviation of Runs from their parent Spec becomes a queryable, analyzable quantity rather than an invisible source of noise. The hierarchy also supports experimental planning workflows: a team can define Process Specs and Material Specs before any physical synthesis occurs, creating a recipe that downstream experimentalists execute as Runs. This explicit separation of design from realization facilitates information handoffs between groups and enables retrospective analysis of how closely experimental outcomes tracked original intentions. This generalizes and matures the \texttt{ideal} concept from the PIF's Composition and Quantity objects, applying it uniformly across all object types.

\subsubsection{Templates as a Flexible Controlled Vocabulary}

A persistent tension in materials data management is between the need for controlled vocabularies and the practical impossibility of imposing a single global ontology on a diverse research community. ``Young's modulus'' in one dataset should recognizably be the same quantity as ``Youngs modulus'' or ``elastic modulus'' in another; however, ``fatigue'' refers to different measurands in a mechanical study and in a battery study.

GEMD addresses this through its template system. \textbf{Attribute Templates} define named, bounded data concepts (e.g., a ``Sintering Temperature'' template with a real value bounded between 200 and 3000~K). By linking attributes to specific Template objects rather than relying on name matching alone, linguistic ambiguities are resolved at the point of data entry, and dataset reconciliation can be performed systematically rather than case by case. \textbf{Object Templates} aggregate Attribute Templates into schemas for specific types of Materials, Processes, or Measurements, and can further restrict inherited bounds to a tighter, application-specific range. They communicate what kinds of attributes are expected for particular tasks and can be used to automatically generate data entry forms.

Critically, template assignment is optional at the object level. Data can be contributed without templates, using free-form name and notes fields, and enriched with template associations retroactively. This ``never turn away data'' principle substantially lowers the barrier to adoption: a new collaborator can begin contributing data immediately without needing to negotiate a global ontology. Comparability between datasets is asserted not by a shared controlled vocabulary imposed from above, but by shared template references negotiated by the community of practice. Two attributes that reference the same Attribute Template are directly comparable; two attributes that merely share a unit dimension may represent conceptually distinct quantities and should be treated with caution.

\subsubsection{Uncertainty Quantification as a First-Class Concern}

Materials measurements are inherently uncertain, yet most data formats store property values as point estimates, discarding distributional information. GEMD treats uncertainty quantification as opt-out rather than opt-in, offering a rich set of Value Types: Normal (Gaussian) distributions parameterized by mean and standard deviation, Uniform distributions for truncation-bounded measurements, and Nominal values for cases where the uncertainty is unknown or unquantifiable. Categorical properties can be expressed as discrete probability distributions over categories.

This design reflects a principled view: uncertainty is a property of measurement, not an optional annotation. By making distributional representations the natural expression for physical quantities, GEMD encourages practitioners to record and propagate measurement uncertainty, which in turn improves the fidelity of downstream statistical analyses and machine learning models trained on GEMD-structured data.

\subsubsection{The Identifier Architecture: UIDs and Tags}

GEMD provides two complementary mechanisms for annotating objects with external or organizational identifiers, each designed for a distinct semantic relationship.

\textbf{Unique Identifiers (UIDs)} assert identity: a \texttt{scope:uid} pair declares that a specific GEMD object \emph{is} the entity known by that identifier in the referenced namespace. This is the appropriate mechanism when the relationship between a GEMD object and an external concept is one-to-one: a particular Material Run \emph{is} laboratory batch 2024-0117; a particular Attribute Template \emph{is} the EMMO concept for temperature. Multiple \texttt{scope:uid} pairs on a single object allow the same entity to be known simultaneously under different naming systems, supporting cross-system lookup without ambiguity.

\textbf{Tags} assert classification membership: a tag on an object declares that the object \emph{belongs to} a named category. Because the same tag may appear on any number of objects, tags are the semantically appropriate vehicle when an external concept maps one-to-many across GEMD objects: a process category code that covers a family of sintering variants, an ASTM standard number that governs a class of measurements, or a project designation shared by an entire experimental campaign. Using a UID for such a relationship would misrepresent it as identity rather than membership.

Tags additionally encode structured hierarchical classification through a \verb|::| path-delimiter convention. 
A tag such as \texttt{synthesis::powder\_processing::sintering} places an object at a specific node in a classification tree, and any system that parses the delimiter can query at any level of that hierarchy---retrieving all objects tagged under \texttt{synthesis} broadly, or only those carrying the specific leaf \texttt{synthesis::powder\_processing::sintering}. 
This taxonomy mechanism requires no external schema; the hierarchy is encoded directly in the string and applies equally to objects and to Templates, allowing entire classes of GEMD objects to be taxonomically positioned alongside individual instances.

Together, UIDs and tags form a two-layer identifier architecture spanning the full range of identifier relationships in materials science: precise entity equivalences across naming systems on one hand, and flexible, multi-valued categorical placements within classification hierarchies on the other. Both layers operate independently of the template system, so classification and vocabulary grounding can be applied to any object regardless of whether it carries a formal template.

\subsubsection{Provenance and Multi-Author Workflows}

The GEMD model explicitly supports the reality that the history of a material often involves multiple individuals, teams, or institutions, each of whom knows only their portion of the workflow. A Process Run or Measurement Run can carry a \textbf{Source} object recording who performed the operation and when. Because the material history is encoded as linked objects rather than a monolithic record, different research groups can contribute their respective segments of a material's history without needing access to or knowledge of other groups' steps. The material and its provenance are assembled by traversing the graph.

This design is particularly valuable for cross-institutional studies, contract research, and supply-chain traceability, where materials change hands repeatedly and full transparency at every step may not be possible or desirable. It also supports blinded measurements, in which a third-party characterization laboratory is deliberately uninformed of the sample's synthesis history.

\subsubsection{GEMD and the FAIR Principles}

The FAIR dimensions introduced in the preceding subsection provide a natural lens for evaluating GEMD's contribution to materials data interoperability.

\textit{Findability} is supported by the two-layer identifier architecture. The UID system allows any object to carry persistent, externally resolvable identifiers across multiple namespaces simultaneously. A single Material Run may carry an internal batch number, a repository accession identifier, and a CAS registry number in parallel~\cite{blaiszik2016mdf}. Tags complement UIDs by providing structured taxonomical placement: the \texttt{::} hierarchical path convention enables objects to be indexed at multiple levels of a classification hierarchy, supporting both broad categorical queries and precise leaf-level retrieval.

\textit{Accessibility} is promoted by GEMD's JSON-based serialization, which is human-readable, widely supported, and free of proprietary dependencies. File link objects allow raw instrument data to be associated with the records that describe them, providing a documented path from processed values back to primary measurements. Source objects attach human-readable provenance, including performer identity and date of execution, directly to process and measurement records, making the chain of custody auditable.

\textit{Interoperability} is the dimension GEMD addresses most directly~\cite{scheffler2022fair,brinson2024fair}. Every GEMD object carries a \texttt{uids} field consisting of an arbitrarily long list of \texttt{scope:uid} pairs. This allows a sintering temperature template to simultaneously carry identifiers referencing EMMO~\cite{emmc2024emmo}, QUDT, and an organization-internal vocabulary, permitting any consuming system that understands one of those vocabularies to resolve the term to a formal semantic definition. The template system provides a mechanism for community groups to define shared semantic structures without requiring universal agreement: two laboratories adopting the same Attribute Template for tensile strength ensure that their data are directly comparable regardless of what local names they use in their own records. The graph structure itself contributes to semantic interoperability by preserving the causal relationships between process steps and material properties, relationships that are systematically destroyed by tabular representations.

\textit{Reusability} is addressed through the combination of the Spec/Run hierarchy, uncertainty-aware value types, and provenance metadata. Specs encode the intended experimental design, providing the context necessary for a data consumer to assess whether data produced under a given protocol are applicable to a new modeling task. The broad range of Value Types ensures that measurement uncertainty is available for downstream statistical analysis rather than discarded at the point of recording. The optional notes fields on all objects preserve the kind of qualitative contextual information, e.g., ``sample showed visible porosity,'' that an expert integrating legacy data would need to assign appropriate statistical weight.

The structured provenance captured by GEMD has direct implications for materials informatics workflows~\cite{ramprasad2017,jain2024ml}. The graph structure naturally generates rich feature representations for synthesis-aware models that incorporate process conditions as inputs alongside composition. The explicit Spec/Run distinction enables training data to be filtered or stratified by process intent versus process outcome, a distinction invisible in flat tabular datasets but potentially significant for model performance. The preservation of measurement uncertainty in GEMD Value Types enables uncertainty-aware modeling approaches, such as Gaussian process regression and Bayesian neural networks, to consume the full distributional information rather than treating every observation as a noiseless point. As the field moves toward high-throughput experimentation and autonomous laboratories~\cite{szymanski2023autonomous}, the ability to accumulate, share, and reason over structured experimental histories will become critical infrastructure, and GEMD provides a strong foundation for that infrastructure.

\namedsubsection{Featurization}
\label{sec:Featurization}
The featurization step transforms the attributes stored in a GEMD material history into features that can be consumed directly by the machine learning algorithms. The Citrine Platform provides a configurable suite of features aligned with the attributes defined in GEMD templates, covering real and categorical properties, chemical formulas, molecular structure, and formulations. During featurization, attributes (properties, conditions, and process parameters) are extracted from specific points in the material processing history to build a feature vector for each material history. Conceptually, this amounts to mapping points on the material history graph to columns in a data table, with the bounds declared in the GEMD templates determining the range of each resulting feature. 

Together, the features extracted from the GEMD graph and other available featurizers capture both the topological information encoded in the GEMD material graph and the domain expertise contributed by users. The most frequently employed classes are described below. 

\subsubsection{Chemical and Molecular Features}
These features represent the stoichiometry and molecular structure of ingredients and materials. Chemical formula features accommodate formulas, such as C$_8$H$_{10}$N$_4$O$_2$, while molecular structure features accommodate the information contained in SMILES or InChI strings, for example CN1C=NC2=C1C(=O)N(C(=O)N2C)C. Featurizers also compute a configurable set of element-level, stoichiometric, or molecular-structure features for individual materials using widely adopted methods from the chemical informatics community, such as the Chemistry Development Kit (CDK)\cite{willighagen_chemistry_2017} and Magpie\cite{goodall2020}, introducing chemically aware information into the ML models.

\subsubsection{Processing Features}
Processing features extract processing and measurement context from a material history. Material properties are often strongly influenced by processing history (e.g., heat treatment) in addition to composition, and processing features enable a repeatable transformation from the parameters and conditions at various points in the material history to unique column locations in the feature vector. This ensures that the full process history is preserved and mitigates a common failure mode in materials ML: pooling data across incompatible processing routes or test conditions merely because they happened to share a column name in a spreadsheet.

\subsubsection{Label Features}
A central design goal of the platform is to enable domain experts to leverage their knowledge of material, formulation, and chemical processes when building models. Label features are one of the primary mechanisms supporting this goal. Labels in GEMD objects mark the user-designated roles that various ingredients play in processing steps (e.g., surfactant or catalyst), injecting domain expertise directly into the feature representation. Labels also facilitate reformulation workflows, such as finding alternatives to toxic or expensive ingredients. During featurization, labels can be extracted as a categorical feature.

\subsubsection{Bespoke Features via Custom Formulae}
Another complementary mechanism for injecting domain knowledge is the ``custom formula'' system, which allows users to define custom features as mathematical functions of other features and latent variables. The resulting expressions are treated as first-class features and included in the model alongside GEMD graph-derived features. For example, F\'eret's Law relates compressive strength of concrete, $f'_c$, to water, cement, and porosity as
$$ f'_c = K \left( \frac{c}{c+w+a} \right)^2 $$
where $c$, $w$, and $a$ are volumes of cement, water, and entrapped air and $K$ is an material-dependent coefficient. In this case, the ratio expression $( c / (c+w+a) )^2$ could be added as a custom descriptor to empower the model with direct access to F\'eret's domain knowledge even without relevant training data.

\subsubsection{Mean-Property and Topological Featurization for Mixtures}
Many industrially relevant materials and products are formed as mixtures of ingredients (for example, polymers with additives), which may themselves be mixtures. The platform can represent these structures either hierarchically, retaining the mixing topology, or in a flattened form. The features derived from a mixture can reflect the relative amounts of its constituent ingredients, the labels those ingredients carry, or both. When many ingredients share a continuous property (e.g.\ molecular weight) and a property of the final mixture depends on that shared property (e.g.\ viscosity), the platform uses a mean-property featurizer to construct ingredient-quantity-weighted features from the constituent properties, handling categorical and missing values gracefully. 

\namedsection{Stage 2: Machine Learning Model Construction}
\namedsubsection{Algorithm Selection}
The choice of ML algorithm is largely driven by the specific materials problem being addressed: the nature of the model inputs, the desired outputs, and the size and structure of the available data\cite{saal2020}. As observed across a wide range of materials informatics case studies, various algorithms have been successfully employed for different problems, and the optimal choice is not always obvious a priori\cite{saal2020}. For continuous composition-dependent design spaces and smaller training datasets typical of materials science, algorithm choice may have a more modest effect on model accuracy than representation or data quality, but can still meaningfully influence performance in specific regimes\cite{saal2020}.

The Citrine Platform addresses this through an automated model selection (AutoML) framework\cite{he2021automl} that removes the burden of algorithm choice from the user. Rather than requiring users to specify a particular ML algorithm, the platform evaluates a set of candidate estimators (with random forests as the default) and automatically selects the best-performing model for the problem at hand\cite{arlot2010}. The procedure is robust by design, gracefully handling the diverse data scenarios encountered in real-world materials development.

Random forests are the primary workhorse of the platform, consistent with their widespread success in materials informatics\cite{saal2020, Ling2017, Breiman2001}. They handle mixed data types (continuous and categorical), are relatively robust to overfitting, perform implicit feature selection, and scale favorably to the high-dimensional input spaces typical of materials problems\cite{Breiman2001, KamHo1998}. Critically for the sequential learning framework described in section \nameref{sec:SL}, the ensemble structure of random forests provides a natural basis for uncertainty quantification. However, the AutoML framework is extensible to additional algorithm families as the platform evolves, and users may also specify particular estimator groups when domain knowledge or prior experience suggests a preferred approach.

\namedsubsection{Well-Calibrated Uncertainty Quantification}
The Citrine Platform implements model-specific uncertainty quantification (UQ) strategies tailored to ensemble tree models, linear ensembles, and Gaussian process regression. Across model classes, the objective is to produce calibrated, heteroscedastic predictive uncertainty estimates suitable for sequential learning (see Section~\nameref{sec:SL}). While different modeling paradigms require distinct formulations, all approaches aim to capture various sources of uncertainty arising from ``small'' and noisy datasets common in materials informatics applications. 

\subsubsection{Random Forests}

Random forests (RFs) are ensembles of decision trees trained on bootstrap resamples of the training dataset. For a test point \( x \), the prediction is the ensemble mean:

\[
\hat{f}(x) = \frac{1}{T} \sum_{j=1}^{T} t_j(x),
\]

where \( T \) is the number of trees and \( t_j(x) \) is the prediction of tree \( j \).

\smallskip\noindent\underline{Bootstrap-Based Variance Estimation}: The random forest uncertainty quantification methodology implemented in the Citrine Platform builds upon the jackknife-based variance estimators developed by Wager, Hastie, and Efron \cite{Wager2014}. This approach was first adapted to data-driven experimental design by Ling et al.~\cite{Ling2017}, who describe it in detail and evaluate it across multiple materials science datasets.

The predicted variance at a test point \( x \) is computed as

\[
\sigma^2(x) =
\sum_{i=1}^{S} \max\left[\sigma_i^2(x), \omega \right]
+ \tilde{\sigma}^2(x),
\]

where:
\begin{itemize}
\item $S$ is the number of training samples,
\item $\sigma_i^2(x)$ is the sample-wise variance contribution associated with training point $i$,
\item $\omega$ is a lower-bound noise threshold,
\item $\tilde{\sigma}^2(x)$ is an explicit bias term.
\end{itemize}

The sample-wise variance term is derived from the jackknife-after-bootstrap (JAB) and infinitesimal jackknife (IJ) estimators, which combine the covariance between bootstrap inclusion counts and tree predictions with leave-one-out corrections and a finite-ensemble Monte Carlo correction. This component captures the epistemic uncertainty arising from finite sampling and the model's sensitivity to perturbations of the empirical distribution.

\smallskip\noindent\underline{Explicit Bias Term}: While jackknife-based variance estimators capture sampling-driven uncertainty, practical scientific and industrial datasets typically exhibit additional error from measurement noise, unobserved variables, or incomplete model specification. To improve calibration in such settings, the Citrine implementation augments the jackknife-derived variance with an explicit, low-capacity bias term:

\[
\sigma^2(x)
=
\sigma_{\text{jackknife}}^2(x)
+
\tilde{\sigma}^2(x).
\]
The bias component is implemented as a shallow decision tree in order to avoid overfitting; its role is only to approximate the residual uncertainty not captured by the jackknife variance. This extension of the classical jackknife formulation was introduced in Ling et al.~\cite{Ling2017}.

\smallskip\noindent\underline{Calibration}: Uncertainty calibration is assessed using normalized residuals:

\[
r_n = \frac{\hat{f}(x_n) - f(x_n)}{\sigma(x_n)}.
\]

Under ideal calibration and independence assumptions, these residuals follow a standard normal distribution. In practice, Ling et al.~\cite{Ling2017} show that the residual distributions are approximately Gaussian with moderately heavy tails, reflecting unmodeled sources of uncertainty. Nevertheless, the jackknife-plus-bias formulation provides substantially improved heteroscedastic calibration compared to constant global error metrics such as the root-mean-square out-of-bag error.

\begin{figure}[ht]
\centering

\begin{subfigure}{1.0\textwidth}
    \centering
    \includegraphics[width=\linewidth]{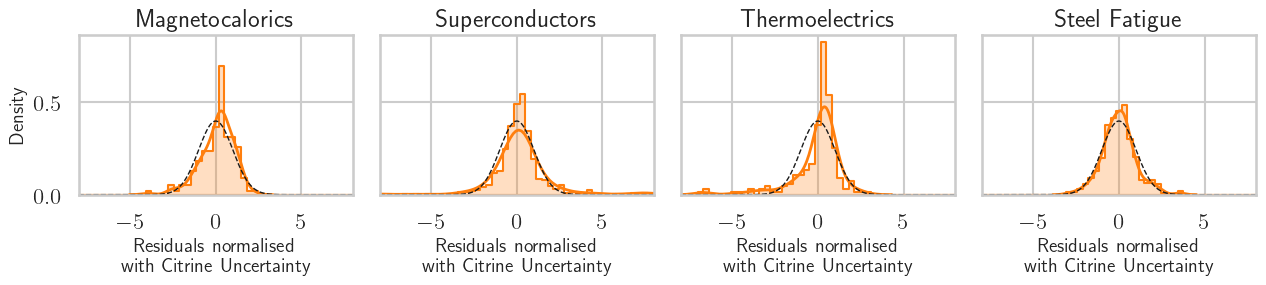}
    \caption{Normalised by Citrine UQ estimate}
\end{subfigure}

\vspace{0.5cm}

\begin{subfigure}{1.0\textwidth}
    \centering
    \includegraphics[width=\linewidth]{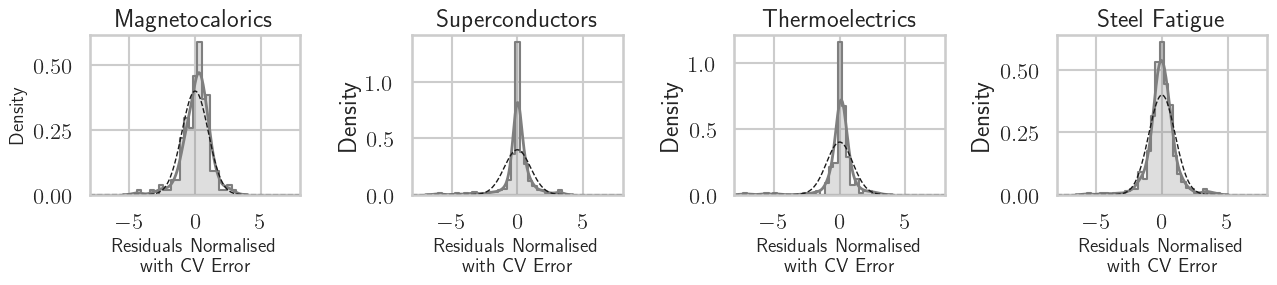}
    \caption{Normalised by out-of-bag error}
\end{subfigure}

\caption{Probability densities of normalized residuals computed via 8-fold cross validation for
each of the 4 cases. The residuals are normalized by (a) the Citrine uncertainty estimate  (b) the root mean square out-of-bag error, which is equivalent to removing the
Jackknife-based uncertainty and using a constant model for the explicit bias. The unit normal
distribution, representing perfectly calibrated uncertainty estimates, is shown for reference
with the dashed line. The continuous line is the KDE density estimate.}
\label{fig:uq_comparison}
\end{figure}

This methodology is implemented in the open-source Lolo library~\cite{lolo} and forms the basis of random forest uncertainty estimation within the Citrine Platform.

\subsubsection{Linear Ensembles}

The Citrine Platform also implements ensembles of linear regression models via bootstrap aggregation. Multiple linear models are trained on independent bootstrap resamples of the dataset,

\[
\hat{f}_j(x) = x^\top \beta_j,
\]

yielding an ensemble of predictions

\[
\hat{f}(x) = \frac{1}{T} \sum_{j=1}^{T} \hat{f}_j(x).
\]

Predictive variance is estimated using the same jackknife-after-bootstrap and infinitesimal jackknife framework described for random forests, optionally augmented by a low-capacity bias term when improved calibration is required. This produces heteroscedastic uncertainty estimates, in contrast to classical linear regression confidence intervals, which assume homoscedastic Gaussian noise.

\subsubsection{Gaussian Process Regression}

Gaussian Processes \cite{Rasmussen2006Gaussian} (GPs) are a common nonparametric modeling framework in materials informatics and the standard choice in Bayesian optimization, but they exhibit some drawbacks relative to Random Forests, including:

\begin{itemize}
\item the need to specify appropriate kernels and learn their hyperparameters
\item sensitivity to heteroscedastic noise
\item learning difficulties in small, sparse, high-dimensional datasets 
\end{itemize}

They nevertheless provide smooth predictions and come equipped with native uncertainty quantification.
Given training inputs \( X \) and targets \( y \), assuming Gaussian observation noise with variance \( \sigma_n^2 \), the posterior predictive distribution at a test point \( x_* \) is Gaussian:

\[
p(f_* \mid x_*, X, y)
=
\mathcal{N}\left(\mu_*(x_*), \sigma_*^2(x_*)\right).
\]

The predictive variance naturally captures both epistemic uncertainty (which increases in regions distant from training data) and aleatoric uncertainty (represented by the noise variance parameter). Unlike ensemble-based methods, GP uncertainty is derived directly from Bayesian posterior inference and requires no resampling-based correction.

\subsubsection{Support Vector Machines (SVM)}

Uncertainty quantification (UQ) for SVM classifiers \cite{platt1999probabilistic, wu2004probability} is derived from the model’s decision function scores. As these scores are not probabilities, they must be transformed to obtain interpretable uncertainty estimates. The transformation depends on whether binary or multiclass classification is performed.
In the binary case, the decision function produces a single score per sample representing the distance from the separating hyperplane. These scores are mapped to a $[0,1]$ interval  using the logistic (sigmoid) function, allowing the outcomes to be interpreted as the probability of belonging to the positive class. Predictions near 0.5 correspond to points close to the decision boundary and therefore higher uncertainty.

In the multi-class setting, the decision function produces one score per class in a one-vs-rest (OVR) format. These scores are converted to probabilities using the softmax function, normalizing the scores across all classes to yield a probability distribution. A prediction is considered uncertain when multiple classes receive similar probabilities.

\namedsubsection{Model Validation via (LOCO) CV}
\subsubsection{Why Random k-Fold CV Falls Short in Materials Discovery}
Cross-validation (CV) is the gold standard approach for quantifying the performance of ML models\cite{saal2020}. In CV, a model is trained on a subset of available data and then evaluated on a held-out partition for which the ground truth is known, probing the model's ability to generalize to unseen examples. The most widely used form is random $k$-fold CV, in which data are randomly divided into $k$ partitions and the model is iteratively trained on $k-1$ folds and tested on the remaining fold. 

Materials data, however, present specific challenges for standard CV: datasets typically exhibit strong clustering; scientists tend to measure many small variations around a few successful parent materials; and discovery often involves extrapolation to chemistries or property regimes not well represented in the training data. As a result, random $k$-fold CV can overestimate model performance relative to what would be observed in a real discovery campaign\cite{saal2020, meredig2018_rsc}.

\subsubsection{Grouping and Stratification in the Platform's CV Framework}
The Citrine Platform implements a configurable CV framework that addresses these challenges through two mechanisms. 

First, a grouping function allows materials to be assigned to groups based on their descriptor values, unique identifiers, or user-defined keys, ensuring that all members of a group are placed in the same fold. This prevents closely related materials (such as compositional variants of the same parent alloy) from being split across training and test sets, which would artificially inflate apparent model accuracy. The effect is analogous to the leave-one-cluster-out (LOCO) approach described by Meredig et al.\cite{meredig2018_rsc}, which was specifically developed to simulate the extrapolative conditions encountered in real materials discovery. 

Second, a stratification layer ensures that folds contain a representative distribution of data types, which is particularly important for sparse or heterogeneous datasets where naive random splitting could produce folds that lack sufficient training examples for certain model outputs. The system first attempts random fold assignment and, if any fold would contain insufficient data for model training, falls back to stratification based on the descriptors relevant to each ML model. Multiple independent trials can be run to reduce variance in the CV estimates. 
  \begin{figure}
      \centering
      \includegraphics[width=\linewidth]{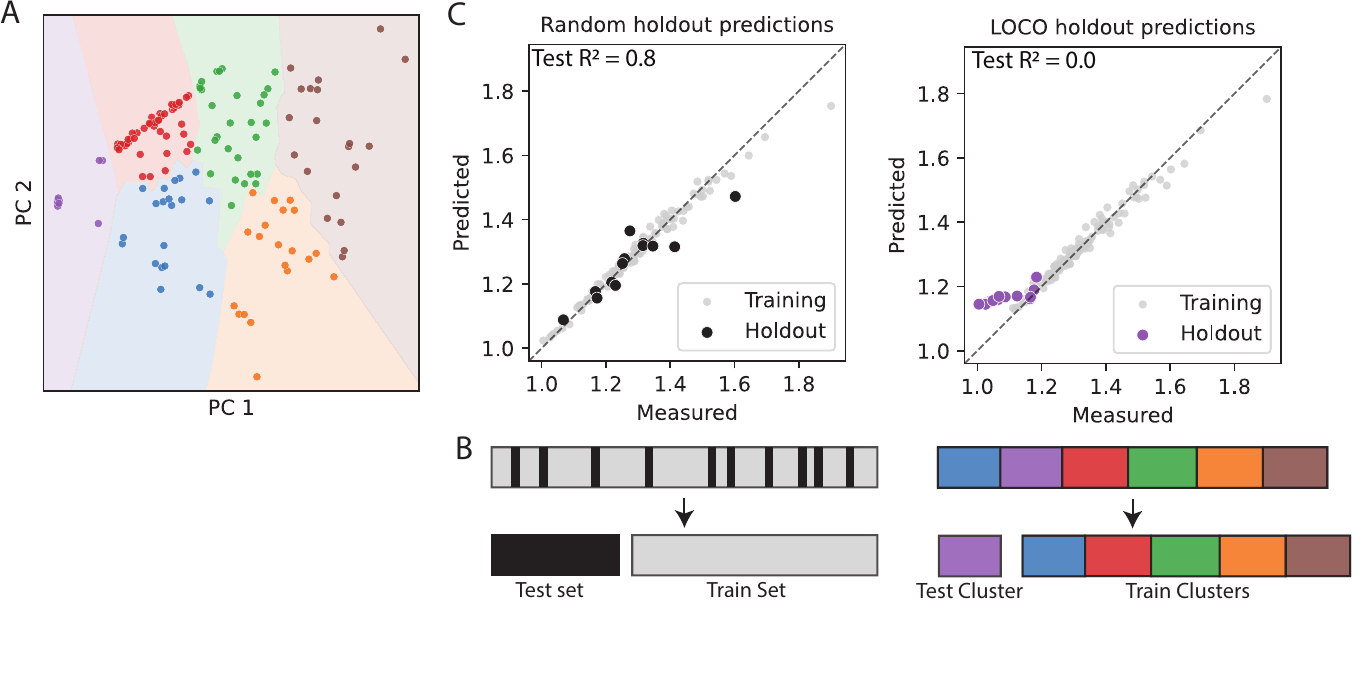}
      \caption{Leave-one-cluster-out cross-validation (LOCO-CV) exposes the optimism of random holdout validation. \textbf{(A)} The dataset partitioned into clusters (colors) by an unsupervised clustering algorithm; for visualization, clusters are shown in principal component space. \textbf{(B)} The two validation strategies compared here. \emph{Left:} random holdout draws the test set (black) at random from across the full dataset, so test points are interspersed among the training data. \emph{Right:} LOCO holds out an entire cluster (purple) as the test set and trains on the remaining clusters, forcing the model to predict in a region of feature space it has never seen. \textbf{(C)} Measured-versus-predicted parity plots for a model evaluated under each strategy (gray, training; colored, holdout; dashed line is the 1:1 ideal). Under random holdout the model interpolates among nearby training points and appears highly accurate ($R^2 = 0.8$), but under LOCO the held-out cluster lies outside the training distribution and predictions collapse to the training mean ($R^2 = 0.0$). The contrast illustrates that random cross-validation can substantially overestimate a model's ability to generalize to new regions of materials space.}
      \label{fig:loco_cv}
  \end{figure}

Together, as demonstrated in Fig.~\ref{fig:loco_cv}, these mechanisms ensure that CV performance metrics provide realistic, actionable estimates of how a model will perform when deployed for materials discovery in a given design space. These accurate performance estimates are crucial because the more the design space diverges from the training data, the more challenging ML-driven discovery will be\cite{saal2020}.

\namedsection{Stage 3: Design Space Definition}
\namedsubsection{Encoding Domain Knowledge Constraints}
Every materials or process optimization campaign starts from existing knowledge.
A formulation chemist knows which ingredients are on hand, which combinations are unstable, and which concentration ranges have produced workable products before.
A process engineer knows the operating limits of each piece of equipment, the tolerance windows the downstream steps require, and the variable relationships the process depends on.
Most of this knowledge lives in lab notebooks, protocol headers, and the working memory of practitioners who filter bad ideas out before they become experiments, rather than in any form an optimizer can read.

The design space formally defines the bounds and limits of the processes, their conditions, and the amounts and types of input ingredients.
Building a design space is where that knowledge becomes explicit and machine-readable. 
Each constraint encodes one piece of what the team already knows, and a more completely specified space leaves less for the optimizer to rediscover from data.

\subsubsection{The Cost of Omission}

Leaving domain knowledge out of the design space does not make the optimization neutral; it makes it uninformed.
A model trained on experimental data learns patterns inside the region that has been explored, but it has no way to tell apart a candidate a practitioner would dismiss on sight from a genuinely promising one.
Without constraints, the optimizer is free to propose candidates in regions the model has never seen.
A well-calibrated surrogate should report high uncertainty there, but in practice predicted uncertainty often stays misleadingly tight: surrogates are calibrated on the observed data and have no mechanism to flag a query that lies outside the regime in which their assumptions hold\cite{Khatamsaz2023}.
The result biases recommendations toward extrapolated edges of the input space, where they are least reliable.

Collecting more data from those regions is the wrong fix.
The right fix is to state, as constraints, what the team already knows: those regions are not worth exploring.

\subsubsection{Constraint Types and What They Encode}

Real design problems need a wider variety of constraint types than most optimization tools support\cite{khatamsaz2022}.
These constraints encode at least three qualitatively different kinds of domain knowledge: \emph{hard infeasibility} (``this candidate cannot be prepared or run,'' e.g.\ a sum-to-one violation, or an ingredient the lab does not stock), \emph{practical exclusion} (``this candidate could be prepared, but no practitioner would intentionally formulate it,'' e.g.\ an ingredient at sub-functional concentration), and \emph{conditional structure} (``some variables are only meaningful given the values of others'').
Each kind needs a different constraint mechanism, and missing any one of them forces workarounds that degrade the search.

\begin{table}[h]
    \centering
    \small
    \begin{tabularx}{\linewidth}{@{} >{\raggedright\arraybackslash}p{2.6cm}
                                       >{\raggedright\arraybackslash}X
                                       >{\raggedright\arraybackslash}X @{}}
    \toprule
    \textbf{Type} & \textbf{Encodes} & \textbf{Example} \\
    \midrule
    Ingredient bound & Range on one ingredient's fraction
        & surfactant $\in [0.5\%, 3\%]$ \\
    Label bound      & Range on a functional-role aggregate
        & total solvent $\in [60\%, 80\%]$ \\
    Dimension bound  & Range on a process or attribute variable
        & cure temperature $\in [80, 120]\,^\circ$C \\
    \addlinespace
    Categorical  & Finite set of allowed discrete values
        & method $\in$ \{injection, casting, extrusion\} \\
    Optionality  & Absent, or within a working range
        & surfactant $\in \{0\} \cup [0.5\%, 3\%]$ \\
    Mixture      & Sum-to-one closure
        & $\sum_i x_i = 1$ \\
    Count        & Cardinality of selected ingredients
        & at most 5 actives; at least 2 fillers \\
    Ratio        & Bounded ratio of weighted sums
        & plasticizer / resin $\in [0.1, 0.3]$ \\
    Conditional  & Validity of one variable depends on another
        & if filler present, surface treatment $\in$ compatible list \\
    \bottomrule
    \end{tabularx}
    \caption{%
        Constraint types supported in the platform's design-space specification.
        The first three rows are the hierarchical levels at which bounds operate; the remaining rows are categories that bounds alone cannot capture.%
    }
    \label{tab:constraint-types}
\end{table}

\textbf{Bounds} are the most familiar constraint, but they operate at three distinct levels in a hierarchical design space.
\textbf{Ingredient fraction bounds} constrain the concentration of a single ingredient: a surfactant must be present between 0.5\% and 3\% by weight; a filler loading cannot exceed 40\%.
\textbf{Label fraction bounds} constrain the aggregate fraction of all ingredients sharing a functional role, so that the total solvent fraction stays between 60\% and 80\% regardless of which specific solvents are selected.
\textbf{Dimension bounds} constrain process conditions and material attributes that sit outside the formulation: a cure temperature must stay within the equipment's operating range, a coating thickness must be achievable by the deposition process, an aging time cannot run shorter than the reaction kinetics allow.
The three levels encode different things: what an individual ingredient can do, what a functional class must contribute, and what the process or material system requires.
Most tools only support the first, which forces practitioners to manually verify the rest.

\textbf{Categorical constraints} restrict a variable to a finite set of allowed values\cite{Zhang2020}.
They show up in two forms depending on where in the design space they sit.
Inside a formulation, the ingredient list itself is the categorical constraint: only materials in the design space's roster can be selected.
This is the definition of what exists, not a constraint layered on top of a continuous space.
A formulation chemist specifies the twenty surfactants available in the lab, and the optimizer cannot propose a twenty-first.
Outside the formulation, process conditions and material attributes are constrained through enumerated dimensions: the processing method must be one of \{injection molding, casting, extrusion\}; the surface treatment is selected from the three qualified options.
Without either form, the optimizer has no concept of inventory, qualification, or prior characterization, and may suggest materials that do not exist or process settings no equipment supports.

\textbf{Optionality} appears in nearly every real formulation problem, but many tools cannot encode it cleanly.
An ingredient is either absent from a candidate, or present at some concentration within a valid range.
The correct encoding follows: if present, the fraction lies between A and B; otherwise it is exactly zero.
Setting bounds to include all fractions from 0 to B wastes effort since it includes the dead zone between 0 and A that no practitioner would intentionally explore.
A surfactant that works between 0.5\% and 3\% is either dosed at a useful loading or left out entirely.
A candidate with 0.1\% surfactant is not a compromise; it is a formulation that wastes an ingredient at a concentration too low to function.
Without proper optionality encoding, the 0-to-A region nevertheless takes up a significant fraction of the search space and attracts recommendations from models that have no way to know it is useless.
Encoding optionality explicitly drops that dead space entirely: the ingredient is either absent or present in its working range.
The result is cleaner training data, sharper model boundaries between ``absent'' and ``effective,'' and candidates the lab can run without guessing whether a trace amount was intentional.

\textbf{Mixture constraints} are not specified by the user; they are an inherent property of formulation design spaces\cite{Cornell2002}.
Any formulation in which the components account for the whole product obeys a sum-to-one requirement, and the platform enforces this automatically at a resolution chosen by the designer (e.g., 1\% increments).
Fractions are discretized at the specified resolution, and every generated candidate sums to exactly one without the user stating it as a separate constraint.
Because the requirement is physical, a system that does not enforce mixture constraints allows the optimizer to propose formulations that seem plausible but are impossible to prepare.
A major challenge in enforcing these constraints is that sum-to-one interacts with bounds on individual components. For example, if three components each have a lower bound of 0.4, their combined minimum already exceeds 1.0 and the design space is infeasible.
Citrine's approach tracks these interactions across the bound hierarchy, ensuring compliance across ingredient bounds, label bounds, count constraints, and the implicit mixture requirement.

\textbf{Ingredient count constraints} specify how many distinct ingredients a formulation may contain --- a minimum, a maximum, or a range, optionally restricted to a subset of ingredients sharing a label such as ``surfactant'' or ``filler.''
These encode practical formulation knowledge: a product with more than five active ingredients is hard to manufacture reproducibly; at least two types of filler are needed to hit the target particle packing; no more than one UV absorber should be present.
Count constraints interact with optionality, since an optional ingredient only contributes to the count when it is present, so they need explicit support rather than a workaround.

\textbf{Ratio constraints} bound relationships between groups of variables.
The ratio of two reactive components must fall within a stoichiometric range; the proportion of plasticizer to total resin must stay within a window that maintains flexibility without sacrificing strength.
The numerator is a weighted combination of one set of ingredients, the denominator a weighted combination of another, and the ratio must fall within specified bounds.
This is more specific than arbitrary inequality support, but it covers the coupling relationships that matter most in formulation chemistry: stoichiometric ratios, component balances, and proportions between functional classes.
Independent bounds on each variable cannot capture these, and learning them from data is impractical, since the model would need many infeasible examples to locate the boundary.

\textbf{Conditional constraints} are the most expressive type and the type most often missing from standard tools\cite{Jenatton2017}.
They specify that the validity of one variable depends on the value of another: if a filler is present, its surface treatment must be chosen from a compatible list; if the process uses injection molding, mold temperature is a relevant input, but if the process is casting, it is not; if the formulation exceeds a certain viscosity, a specific dispersant must be included.
This kind of if-then structure shows up wherever components or process steps are themselves optional or categorical, which is to say almost everywhere.

Even without fully general conditional-constraint support, many practical dependencies can still be represented through combinations of labels, bounds, optionality, and ingredient count constraints.
These mechanisms capture common formulation rules such as mutual exclusion, minimum functional coverage, or restrictions on how many materials from a category may appear together.
In practice, this allows a large fraction of real design-space structure to be expressed without requiring arbitrary if-then logic.

The standard workaround in tools without either conditional constraints or this label-count mechanism is to include all variables for every candidate and use a sentinel value (zero, ``N/A,'' or a flag) to mark variables that do not apply.
The model then has to learn the conditional structure from data, inferring (for example) that mold-temperature predictions should be ignored whenever the casting flag is set, when the structure could have been stated directly.
The generated recipes also mix meaningful settings with sentinel placeholders, leaving the lab to decide case by case which fields to act on and which to ignore.
That manual triage step defeats the purpose of automated recommendation.

\subsubsection{From Individual Constraints to a Coherent Space}

The design space draws its power from the combination of constraints, not from any single one.
A formulation that satisfies bounds on every ingredient individually can still violate a mixture constraint or an ingredient count limit.
A candidate that passes all scalar constraints can still trip a conditional dependency that renders one of its variables meaningless.
Evaluating constraints jointly, rather than independently, is what produces candidates that are coherent as a whole rather than merely valid along each dimension\cite{Rossi2006}.

Joint evaluation also enables something individual constraints cannot: detecting infeasibility before any sampling or optimization runs.
If the minimum fractions of required ingredients already exceed the maximum total fraction, the space has no valid candidates.
Catching that at specification time, rather than after a failed optimization run, saves real time.
The design space specification is the first place the consistency of domain knowledge can be checked, and the earlier an inconsistency is caught, the less it costs.

\namedsubsection{Design Space Enumeration and Generation}
A design space specifies what is allowed; searching it produces specific experiments to evaluate.
How candidate generation works, and how it scales, sets a hard ceiling on the problem sizes the optimization can handle.

\subsubsection{When Enumeration Is Sufficient}

The simplest generation strategy is enumeration: list every valid candidate and let the optimizer pick.
This works well for small, discrete spaces such as a fixed catalog of known materials, a short list of candidate formulations from a previous study, or a set of process variants under evaluation.
Enumeration requires no sampling, introduces no approximation, and guarantees that every candidate in the search is exactly what it appears to be.

Scale is the catch.
Enumeration only works while the candidate set is small enough to hold in memory and score directly, and that ceiling is reached fast.

\subsubsection{The Combinatorial Explosion Problem}

\begin{figure}[h]
    \centering
    \includegraphics[width=0.8\linewidth]{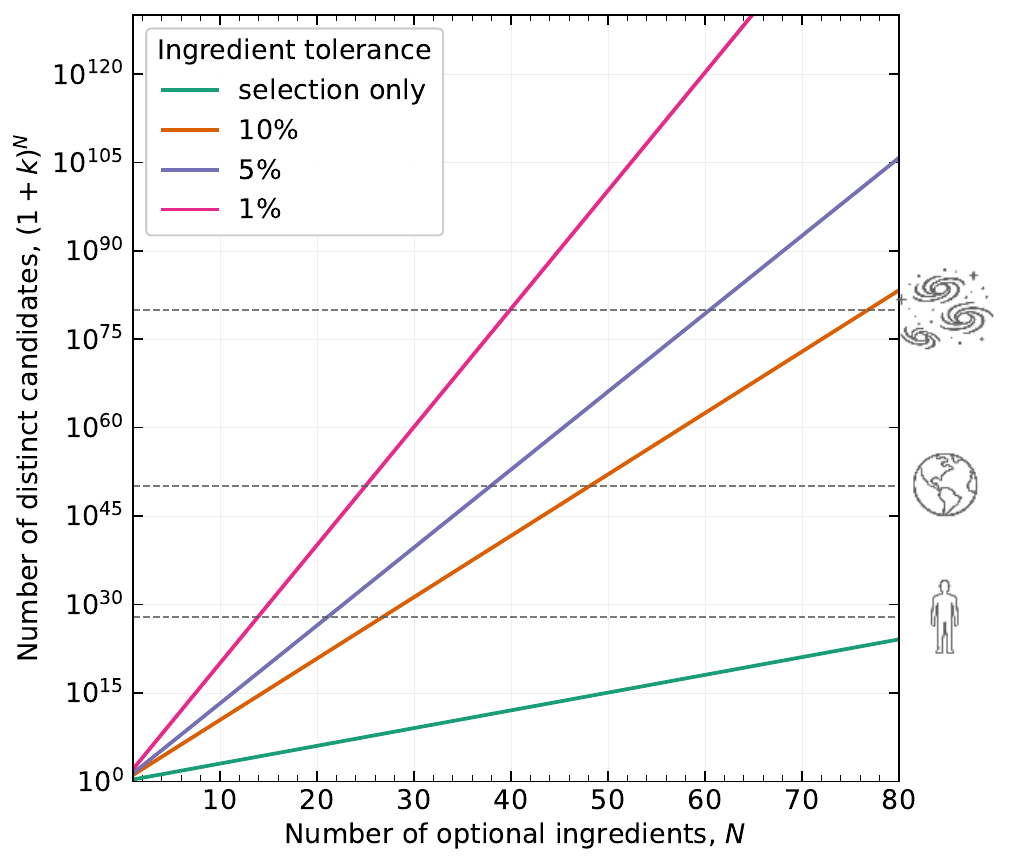}
    \caption{%
        Growth of the formulation design space as a function of the number of optional ingredients $N$, for four ingredient-tolerance settings (legend, upper left).
        Each setting corresponds to a practical lab dosing tolerance: ``selection only'' is $k=1$ (in or out), $10\%$ tolerance is $k=10$, $5\%$ is $k=20$, and $1\%$ is $k=100$.
        The total candidate count scales as $\sum_{i=0}^{N} \binom{N}{i} k^i = (1+k)^N$, far outstripping any fixed experimental budget within a handful of ingredients.
        Dashed horizontal lines give the number of atoms in a human body ($\sim 10^{27}$), the Earth ($\sim 10^{50}$), and the observable universe ($\sim 10^{80}$), as physical anchors for the count; icons at the right edge mark each level.%
    }
    \label{fig:combinatorial-explosion}
\end{figure}

Consider a formulation design space with $N$ candidate ingredients, each optional.
When an ingredient is present, its fraction is chosen from $k$ discrete values, where $k$ corresponds to the dosing tolerance the lab can hit: $k=10$ means $10\%$ control, $k=20$ means $5\%$ control, and $k=100$ means $1\%$ control.
The case $k=1$ is selection-only, where each ingredient is either included at one fixed level or absent; we use it as a lower bound.
Counting all formulations is then a two-step process.
First, the \emph{selection} step: each of $N$ ingredients is either in or out, giving $2^N$ possible subsets.
Second, the \emph{concentration} step: each of the $i$ ingredients in a chosen subset takes one of $k$ values, giving $k^i$ options.
Summing over all subset sizes,
\begin{equation}
    \sum_{i=0}^{N} \underbrace{\binom{N}{i}}_{\text{subset selection: } 2^N} \cdot \underbrace{k^i}_{\text{concentrations}} \;=\; (1+k)^N.
\end{equation}
For $N=20$, the selection step alone yields $2^{20} \approx 10^6$ subsets before any concentration is chosen; at $k=10$ the full count is $11^{20} \approx 7 \times 10^{20}$ (Figure~\ref{fig:combinatorial-explosion}).
Add a second categorical variable, say a choice of processing method, and the count multiplies again.
Add a third, and a fourth, and enumeration is no longer an option.

This is the combinatorial explosion that rules out enumeration in most real design problems\cite{Wu2024}, and it shows up common cases: formulations with more than a handful of optional ingredients, processs with more than a few discrete settings, or hierarchical spaces with branching structure.
Candidate counts outpace experimental budgets by many orders of magnitude.

The size of the design space matters for reasons beyond search strategy.
A space with ten valid candidates and a space with ten billion call for different experiment-selection schemes, different expectations about coverage, and different interpretations of what the training data represents.
Estimating the size of a constrained design space requires reasoning about mixture constraints, count constraints, label fractions, and optionality together, since they interact.

\subsubsection{Sampling as an Alternative to Enumeration}

When a design space is too large to enumerate, the platform generates candidates by sampling.
It draws a set of candidates that satisfies all constraints and spreads across the feasible region, and the optimizer selects among them based on predicted performance.

Sampling under constraints is much harder than sampling without them.
Drawing a random point and checking whether it satisfies the constraints, known as rejection sampling, only works when the feasible region is a large fraction of the total space\cite{Wu2024}.
For formulations with tight mixture constraints, count limits, and ingredient bounds, the feasible region can be a tiny sliver of the naive sampling domain, and most random draws are rejected.
Rejection sampling gets less efficient exactly as constraints get tighter, which is the regime where good constraint specification matters most.

The platform handles this with constraint-aware samplers that incorporate the full constraint set directly into candidate generation rather than filtering invalid candidates afterward\cite{Rossi2006,GriffithsHernandezLobato2020}.
Different sampling strategies are appropriate for different classes of constrained problems, including continuous mixture spaces, hierarchical conditional structures, and objective-driven optimization.
In every case, constraints enter the generative process directly, so generated candidates are feasible by construction and sampling effort is concentrated in regions of the design space that matter.

The sampler also enforces diversity, ensuring that each round of experiments covers distinct regions of the feasible space rather than clustering around a single promising area.

\subsubsection{Multi-Objective Optimization}

Few materials campaigns optimize a single property\cite{khatamsaz2022}.
A structural adhesive must be both strong and tough; pushing strength alone usually costs toughness.
A coating must be durable, flexible, and manufacturable at acceptable viscosity, and gains in one property typically erode another.
Reasoning about multiple objectives at once is a prerequisite for finding useful candidates.

The platform handles multi-objective optimization through the acquisition function, which scores each candidate using the model's predicted properties and uncertainties; higher-scoring candidates are preferred.
Different acquisition functions encode different stances on exploration versus exploitation, and on the trade-offs between competing objectives.
Relative priorities are expressed through baselines and the structure of the acquisition function itself~\cite{Balandat2020}, both of which can be updated as the campaign progresses and the team's view of the trade-off surface sharpens.
Section~\nameref{sec:acquisition} describes the acquisition functions available in the platform and the regimes where each performs well.

\subsubsection{Connecting Generation to the Optimization Loop}

Candidate generation, constraint satisfaction, and acquisition-function scoring are not independent steps; together they form the sequential learning loop described in Section~\nameref{sec:SL}.

The loop's efficiency tracks the quality of the design space at every stage.
A space that is too broad generates candidates the lab cannot run.
A space encoding invalid conditional dependencies generates candidates that need manual correction.
A space whose size cannot be estimated leaves the team unable to say how much of it has been explored.
Design space specification is ongoing work that runs alongside the optimization, updated as experimental results show which assumptions held and which need revision.

\namedsection{Stage 4: Optimization within the Design Space}
\namedsubsection{The Sequential Learning Framework}
\label{sec:SL}
To efficiently search the design space for the best materials, the Citrine Platform employs an iterative approach known as sequential learning (SL)~\cite{Ling2017, Cohn1996}. 
In this framework, the platform repeatedly recommends promising candidate experiments, users obtain validation data through measurement or simulation, and the platform retrains the machine learning models on the resulting data. This closed-loop process enables rapid convergence toward optimal materials even when starting from datasets that are sparse or do not sample the target design space.

The SL process starts by training a model on an initial set of training data (Figure ~\ref{fig:SL_fig}). In addition to learning to predict properties of new materials, the model estimates the uncertainties of those predictions.
The platform then samples the design space to generate a list of candidates.
The trained model takes in the list of candidates to produce predicted properties and estimated uncertainties, all of which are used to score the candidates using an acquisition function~\cite{Shahriari2016}. The acquisition function decides how to balance exploration (choosing candidates that best improve model performance) and exploitation (choosing candidates that best fulfill the target optimization criteria) when scoring candidates. Incorporating their expertise, users can then choose from the top-scoring candidates to synthesize and characterize new materials. The model is then retrained with the new data and the process iterates until the design targets are achieved.

\namedsubsection{Acquisition Functions}
\label{sec:acquisition}
The acquisition function determines how model predictions and uncertainties are combined to score candidate experiments, balancing the fundamental trade-off between exploitation (selecting candidates predicted to perform well) and exploration (selecting candidates where the model is uncertain and new information would be most valuable). The Citrine Platform implements three acquisition functions, each encoding a different stance on this trade-off. Because the platform's internal naming conventions evolved independently from the broader Bayesian optimization (BO) literature, we provide here both the platform terminology and the corresponding standard names to facilitate cross-referencing.

\begin{figure}
    \centering
    \includegraphics[width=\linewidth]{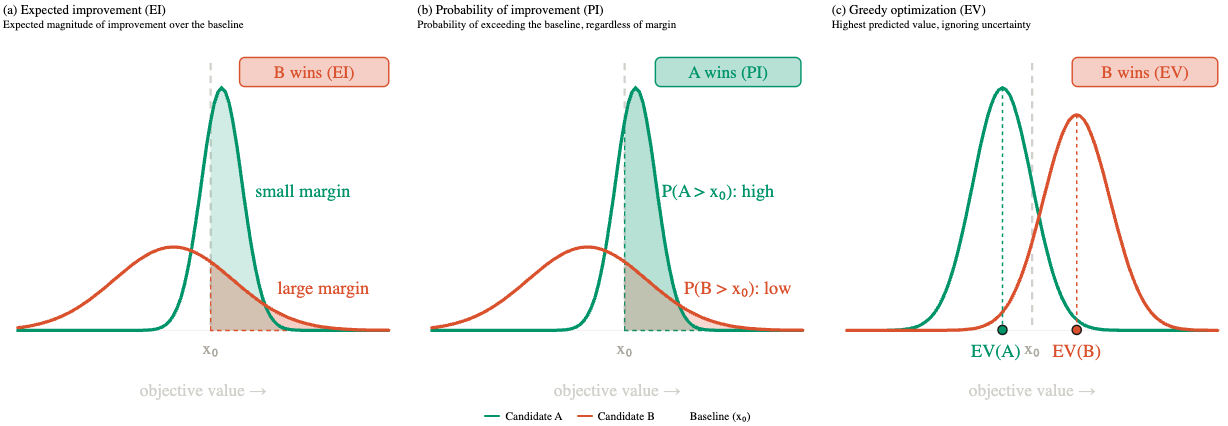}
    \caption{Comparison of three acquisition functions applied to candidate predictive distributions. In panels (a) and (b), the same two candidates A and B are evaluated: A is a narrow distribution centered just above the baseline $x_0$, while B is a broad distribution centered below $x_0$ with a long right tail. EI (a) favors B because the expected magnitude of improvement (the shaded area weighted by distance from $x_0$) is larger, despite B's lower probability of exceeding the baseline. PI (b) favors A because a greater fraction of A's probability mass lies above $x_0$, even though any improvement is marginal. This contrast illustrates how EI up-weights high-uncertainty candidates with large potential payoffs, while PI favors candidates likely to yield small gains. Panel (c) shows greedy optimization, which selects purely on predicted mean, ignoring both uncertainty and the baseline.}
    \label{fig:acquisition}
\end{figure}

\textbf{Expected Improvement} (EI), referred to internally as Maximum Expected Improvement (MEI), is the classical acquisition function from the BO literature\cite{Mockus1978, Jones1998, Shahriari2016}. For a single objective, it computes the expected magnitude of improvement over the current best-known value (the baseline), defined as $\text{EI}(\mathbf{x}) = \int \max(x - x_0, 0) \cdot p(x) \, dx$, where $x_0$ is the baseline and $p(x)$ is the predictive distribution. When the predictive distribution is Gaussian, this integral admits a well-known closed-form solution in terms of the standard normal PDF and CDF\cite{Shahriari2016}; for non-Gaussian predictive distributions, the platform employs numerical quadrature. EI naturally balances exploitation and exploration: candidates with high predicted mean and candidates with high uncertainty both receive elevated scores, since either can contribute to a large expected improvement. EI is currently limited to single-objective problems. In an SL benchmarking study by Borg et al.\cite{borg2023}, EI emerged as a robust default choice across diverse materials design problems, although no single acquisition function dominated in all scenarios.

While EI quantifies the expected magnitude of improvement, rewarding candidates in proportion to how much better they could be, Probability of Improvement takes a complementary approach by quantifying the likelihood of any improvement at all, regardless of its size.

\textbf{Probability of Improvement} (PI), referred to internally as Maximum Likelihood of Improvement (MLI), scores each candidate by the probability that it simultaneously improves upon the baseline values for all objectives and satisfies all constraints. PI traces to one of the earliest formulations of model-based sequential optimization\cite{Kushner1964}. PI is computed as the joint likelihood over a set of inequality constraints: the original user-defined constraints plus additional constraints requiring improvement over each baseline. This formulation naturally extends to multiple objectives without requiring Pareto scalarization; a candidate scores highly only if it is likely to improve on every objective at once. In early benchmarking by Ling et al.\cite{Ling2017}, PI (as MLI) consistently achieved the highest performance across four materials science test cases spanning magnetocalorics, superconductors, thermoelectrics, and steel fatigue strength. Because PI is indifferent to the magnitude of improvement, it can favor "safe" candidates with high probability of marginal gains, whereas EI naturally up-weights candidates with larger potential payoffs even when improvement is less certain

\textbf{Greedy optimization} (pure exploitation), referred to internally as Maximum Expected Value (MEV), selects the candidate with the highest predicted value of the objective, weighted by the probability of constraint satisfaction. This is equivalent to both Upper Confidence Bound (UCB) with an exploration parameter $\kappa = 0$\cite{Srinivas2010} and minimization of simple regret\cite{Rohr2020}. Because it ignores prediction uncertainty entirely, this strategy can converge quickly when the model is already accurate but risks becoming trapped in local optima when the model is poorly calibrated or the design space is underexplored. The Citrine Platform implementation of greedy optimization supports multiple simultaneous objectives, summing their expected values and incorporating constraints via a hinge function on the joint probability of constraint satisfaction.

A key implementation detail across all three acquisition functions is that constraint satisfaction is treated as a first-class component of the score rather than a hard filter. Each candidate's score is modulated by the joint probability that all user-defined constraints are satisfied, ensuring that the optimizer naturally steers toward feasible regions of the design space while still considering candidates near constraint boundaries where predictions are uncertain. Borg et al.\cite{borg2023} provide a systematic framework and set of metrics, including discovery acceleration factor, discovery yield, and discovery probability, for evaluating acquisition function performance across different target property ranges and iteration budgets, and find that the optimal choice depends on the specifics of the design problem at hand. The Citrine Platform allows users to select and switch between strategies as their campaigns progress, informed by these diagnostic tools.

\namedsubsection{Efficiency and Scalability}
Efficiency in the sequential learning context has two distinct and largely independent meanings. The first is \textit{discovery efficiency}: the number of experiments required to identify high-performing materials, which determines how quickly a materials design campaign can reach its objective. The second is \textit{computational efficiency}: the wall-clock time required to complete each SL iteration, which determines whether the platform can keep pace with real experimental workflows and scale to industrially relevant design spaces. We address each in turn.

\subsubsection{Discovery Efficiency and the Iteration Budget}

The core claim of sequential learning is that uncertainty-guided candidate selection reduces the number of experiments needed to find high-performing materials relative to uninformed strategies. Borg et al.~\cite{borg2023} quantify this advantage through the Discovery Acceleration Factor (DAF$_n$), defined as the ratio of the number of SL iterations required by random search to find $n$ target-range compounds to the number required by a given acquisition function. Using retrospective simulations on band gap and thermoelectric materials datasets with $n_{\mathrm{iter}} = 100$ iterations and $n_{\mathrm{batch}} = 1$ candidate selected per iteration, they find that EI and EV (Expected Value, greedy exploitation) consistently outperform random search across nearly all target deciles, with the advantage growing substantially as the number of targets sought increases. For band gap targets in the extreme deciles, DAF$_5$(EI) reaches 5.6 and 5.0 for the 1st and 10th deciles respectively, indicating that EI identifies five target-range compounds approximately four to five times faster than random search under these conditions~\cite{borg2023}. For intermediate-value targets (5th and 6th deciles), the advantage narrows to DAF$_5 \approx 3.0$, reflecting the greater density of candidates near the distribution center and the correspondingly weaker informational value of uncertainty-guided selection in that regime.

An important practical implication of these results is that the iteration budget --- the number of SL cycles a campaign can sustain --- has a strong effect on the achievable Discovery Probability (DP$_i$): the likelihood that a search has identified a target-range material by iteration $i$~\cite{borg2023}. In early iterations, acquisition functions that incorporate uncertainty (EI) tend to behave similarly to exploratory strategies as the model is poorly calibrated over much of the design space; the advantage of uncertainty-aware selection compounds in later iterations as the model accumulates targeted knowledge of the property distribution tails. Campaigns with fewer than approximately ten iterations see modest SL benefit over random selection regardless of acquisition function choice; the full acceleration potential requires iteration budgets commensurate with the rarity of the target. Roughly one order of magnitude more iterations than the expected number of target compounds in the design space provides a practical guideline. Users should therefore plan campaigns with a realistic assessment of experimental throughput against this benchmark.


\subsubsection{Batch Selection and Experimental Parallelism}

The simulations in Borg et al.~\cite{borg2023} and the analysis above assume $n_{\mathrm{batch}} = 1$: a single candidate is selected and measured per SL iteration. In practice, experimental throughput often permits (or requires) evaluating multiple candidates in parallel before the model is retrained. When $n_{\mathrm{batch}} > 1$, the acquisition function must select a batch of candidates jointly rather than sequentially, and the theoretical optimality guarantees of single-point acquisition functions no longer strictly apply.

Several principled strategies for simultaneous batch selection have been developed in the Bayesian optimization literature. The most theoretically rigorous is the batch or $q$-point Expected Improvement ($q$-EI), which evaluates the joint expected improvement of a set of $q$ candidates simultaneously by integrating over their correlated predictive distributions~\cite{Balandat2020}. Because the joint integral is intractable in closed form, Monte Carlo approximations are required, making $q$-EI substantially more expensive to compute than its single-point counterpart; practical implementations such as BoTorch~\cite{Balandat2020} address this through differentiable sample-path estimators and GPU-accelerated quadrature. A computationally lighter alternative is the Kriging Believer (KB) or Constant Liar strategy~\cite{ginsbourger2010}, which builds the batch sequentially: after each candidate is selected by the single-point acquisition function, its unobserved property value is hallucinated as the model's current prediction (Kriging Believer) or as a fixed constant such as the current best-known value, and the model is temporarily updated before the next candidate is chosen. This transforms the batch problem into a sequence of single-point problems at the cost of one synthetic model update per additional batch member. Local penalization~\cite{gonzalez2016} takes a complementary approach, discouraging selection of candidates that are spatially proximate to already-chosen points in feature space by adding a smoothly decaying penalty term to the acquisition scores, naturally promoting batch diversity without requiring a model refit. Thompson Sampling offers yet another route: independently sampling a realization of the property function from the posterior distribution for each batch slot and selecting the corresponding argmax produces batches that are diverse by construction while remaining statistically consistent with the model~\cite{Shahriari2016}.

The Citrine Platform's current implementation returns a ranked list of top-scoring candidates and invites the user to select the final experimental batch from among them. This design is deliberate rather than a limitation. In the low-data regimes for which the platform is specifically optimized --- campaigns with tens to low hundreds of training points --- the model has not yet had sufficient data to encode the full structure of the design space from the data alone. At these data volumes, a substantial share of the total information relevant to candidate selection is found in domain knowledge that a practitioner holds implicitly, e.g., awareness of a supplier's current material inventory, intuitions about which synthesis routes are feasible given available equipment, prior beliefs about failure modes that have not yet manifested in the data. Algorithmic batch diversity methods operate strictly within the noise floor of the current model; they cannot access knowledge that has not yet crossed the threshold from human intuition into experimental measurement. By surfacing a curated shortlist and delegating the final selection to the user, the platform allows this tacit knowledge to shape the batch in ways that a fully automated strategy would systematically overlook. As the campaign matures and the model's coverage of the design space increases, the relative value of algorithmic diversity versus human curation shifts, and tighter automated batch construction becomes more warranted.


\subsubsection{Computational Scaling with Design Space Size}

Each SL iteration involves three computationally distinct steps: featurizing the candidate pool, training the machine learning model on the updated training set, and scoring all candidates with the acquisition function. These steps scale differently with the relevant problem dimensions.

Featurization of the candidate pool is embarrassingly parallel and scales linearly with the number of candidates $N_{\mathrm{cand}}$. For composition-based features of the kind used throughout this platform (\hyperref[sec:Featurization]{Featurization}), per-candidate featurization is a lightweight operation, and featurizing pools of $10^5$--$10^6$ candidates is tractable on a single compute node without specialized hardware. Model training with random forests scales as $\mathcal{O}(n_{\mathrm{trees}} \cdot n_{\mathrm{train}} \cdot n_{\mathrm{feat}} \cdot \log n_{\mathrm{train}})$ in the size of the training set $n_{\mathrm{train}}$, which grows incrementally with each SL iteration and remains small relative to $N_{\mathrm{cand}}$ throughout any realistic campaign. Because model training depends on the training set size rather than the candidate pool size, it does not become a bottleneck as the design space grows. Jackknife-based uncertainty estimation~\cite{Wager2014}, which requires holding out each training point in turn during ensemble prediction, adds a constant multiplicative factor to training cost but does not change its scaling behavior.

Acquisition function scoring requires one model forward pass per candidate, making it $\mathcal{O}(N_{\mathrm{cand}})$ in the candidate pool size. For Expected Improvement with a Gaussian predictive distribution, the closed-form solution~\cite{Shahriari2016} reduces the per-candidate cost to an inexpensive arithmetic operation once the predictive mean and variance are available, so the bottleneck for large design spaces is typically the model inference step rather than the acquisition computation itself.



The combination of linear acquisition scoring and training-set-limited model training means that the SL pipeline's computational cost is dominated by featurization and inference at large $N_{\mathrm{cand}}$, both of which are highly amenable to parallelization. The design spaces accessible to the platform therefore extend well beyond the $N_{\mathrm{cand}} = 10^3$-scale candidate pools studied by Borg et al.~\cite{borg2023} to the very large pools that arise from combinatorial enumeration of multi-component alloy or formulation spaces (Figure~\ref{fig:combinatorial-explosion}). Enabling this scale without sacrificing per-iteration latency is a prerequisite for applying SL to the full complexity of real industrial materials design problems.

\namedsection{Applications}

Active learning is now a standard application of machine learning models within the materials informatics community. A review of published efforts whose predictions were verified experimentally or computationally~\cite{saal2020} details a large variety of materials classes and properties, training dataset sizes, design space sizes, and ML algorithms. In this section, we summarize published applications of FUELS-based sequential learning specifically, where quantified uncertainty in the acquisition functions is a critical feature in enabling the rapid development of novel materials and manufacturing optimization. 

\subsubsection{High-Mobility Organic Hole-Transport Semiconductors}
Antono et al.~\cite{Antono2020} applied sequential learning to the discovery of high-mobility organic hole-transport semiconductors, coupling a random-forest surrogate model to expensive quantum simulations. From an initial dataset of 32 DFT-calculated hole mobilities, the authors trained the surrogate model with uncertainty quantification on 229 scalar molecular descriptors derived from SMILES strings. The FUELS active-learning framework iteratively selected candidates from large enumerated design spaces, including a filtered 130{,}000-molecule subset of the Harvard Clean Energy Project database and a custom library of polycyclic aromatic molecules with heteroatom substitutions, using the FUELS active-learning framework. Three acquisition functions were applied in sequence: maximum uncertainty, maximum expected improvement, and maximum likelihood of improvement (MLI) acquisition strategies. Selected candidates were evaluated via combined DFT and molecular dynamics simulations, with charge mobility in amorphous phases was computed with DFT from MD-derived bulk morphologies. Over 60 sequential learning cycles involving 165 new DFT/MD evaluations, the workflow screened more than a million candidate structures and ultimately identified a fused thioacene molecule with a predicted hole mobility of $10^{-1.86}$~cm$^2$/(V s), exceeding the best value in the initial training data ($10^{-1.96}$~cm$^2$/(V s)). The study demonstrates that sequential learning can efficiently navigate large chemical design spaces and enable extrapolative discovery of higher-performance organic semiconductors from limited initial data. 

\subsubsection{Autonomous Palladium Nanoparticle Synthesis}
Fong et al.~\cite{Fong2021} applied sequential learning to process optimization, identifying nanoparticle synthesis conditions in a data-limited regime. The specific application was colloidal synthesis of palladium nanoparticles, and integrating automated flow-reactor synthesis with \text{in situ} small-angle x-ray scattering (SAXS) to measure nanoparticle size, polydispersity, and yield in real time. A  16-experiment cold start was used to train a random forest model to predict synthesis outcomes across a multidimensional parameter space that included precursor concentration, reducing-agent concentration, temperature, and residence time in the flow reactor. In each round, the model proposed new experiments that were executed by the automated hardware and added to the dataset, forming an autonomous closed-loop optimization cycle. The autonomous workflow identified conditions producing monodisperse Pd nanoparticles with targeted diameters in the $\sim$1–-10 nm range and simultaneously revealed relationships between synthesis parameters and particle size distributions. The ML-guided experimental loop converged toward optimal conditions on the order of tens of experiments, substantially fewer than the hundreds typically required by traditional trial-and-error exploration, illustrating the potential of autonomous experimentation combined with probabilistic machine learning to accelerate materials synthesis discovery and establish predictive design rules for nanoparticle growth.

\subsubsection{FUELS Benchmark Case Studies}
The original publication of the FUELS framework~\cite{Ling2017} evaluated sequential learning performance by simulating a design task on four materials datasets: magnetocalorics, superconductors, thermoelectrics, and steel fatigue strength. Each was formulated as the task of identifying the optimal candidate from a finite set of previously characterized materials while minimizing the number of required evaluations. In the magnetocaloric case, the algorithm searched among 167 candidate compounds for the material with the largest magnetic deformation and located the optimal candidate in 47$\pm$3 evaluations using the MLI strategy, compared with 84 evaluations expected from random sampling. For superconductors, the search space contained 546 candidate materials with measured critical temperatures, including the best-performing compound HgBa$_2$Ca$_2$Cu$_3$O$_8$; sequential learning identified this optimum in roughly 52$\pm$5 to 98$\pm$12 evaluations depending on the strategy, whereas random search required 273 evaluations on average. In the thermoelectrics test case, involving 195 materials with measured ZT at 300~K, sequential learning located the optimal candidate in 29–-37 evaluations, compared with 98 evaluations for random sampling, more than halving experimental effort. Finally, in the steel fatigue strength problem, combining 437 candidate combinations of alloy composition and processing parameters, the optimal candidate was identified in 24$\pm$2 evaluations using the MLI strategy, while random guessing required 219 evaluations on average. Across these case studies, the sequential learning strategies consistently reduced the number of required measurements by factors of roughly 2–-9$\times$ relative to random exploration. Performance was also compared with the Bayesian optimization framework COMBO, which achieved comparable results, indicating that the random-forest–based sequential learning approach can match state-of-the-art Bayesian optimization while remaining computationally efficient in high-dimensional materials design spaces. 

\subsubsection{Sequential Learning Discovery Metrics}
Borg et al.~\cite{borg2023} performed a simulated sequential learning study that investigated SL efficacy under different scenarios and questioned the common assumption that predictive accuracy translates directly to discovery performance. Simulated sequential learning campaigns were performed on datasets of thermoelectric properties drawn from the Starrydata2 database~\cite{katsura2019starrydata} and experimental band gaps from the Matbench benchmark suite~\cite{dunn2020matbench}, originally compiled by Zhuo et al.~\cite{zhuo2018bandgap}. The study demonstrates that conventional static model metrics such as RMSE or R$^2$, which measure average prediction error across a dataset, are often poorly correlated with a model’s ability to identify high-performing materials during sequential discovery. Instead, the success of a machine-learning-guided search depends strongly on factors such as the target region of the property distribution (e.g., whether the goal is to discover top-1\% versus top-10\% materials), the inclusion of predictive uncertainty in the acquisition function, whether the objective is to discover a single optimal material or multiple high-performing candidates, and the total number of sequential learning iterations allowed. To more appropriately quantify discovery performance, the authors introduce evaluation metrics tailored to sequential learning workflows, including Discovery Yield (the number of high-performing materials discovered during the search) and Discovery Probability (the likelihood that a search has discovered a high-performing material at a given iteration). Through these analyses, the work highlights that models with relatively high prediction errors can still perform well in guiding materials discovery if they correctly rank promising regions of the design space, emphasizing that the evaluation of machine-learning models for materials discovery should focus on their dynamic performance in closed-loop optimization rather than solely on traditional regression accuracy metrics.

\namedsection{Conclusions}
Across this overview, the Citrine Platform has been described as a sequence of four cooperating stages rather than a single system, and that organization is intentional. At the data layer, GEMD treats process history, measurement uncertainty, and provenance as first-class features of every record, and its Template--Spec--Run hierarchy lets a community grow shared vocabularies without committing to a single global ontology up front. On top of that data layer, the model-construction pipeline insists on well-calibrated uncertainty (now extended to multivariate prediction intervals when objectives are correlated~\cite{folie2023mlst}) and validates models the way they are actually used: on extrapolative splits and dynamic discovery metrics, not on random held-out sets~\cite{meredig2018_rsc,borg2023}. Where most published screening pipelines stop once candidates have been ranked, the platform pushes further, encoding compositional, physical, processing, and economic constraints directly into the design space so that candidates that cannot be made or shipped are never proposed in the first place. The sequential learning stage then closes the loop: surrogate models with calibrated uncertainty drive acquisition functions that pick the next experiment, and the platform navigates large constrained spaces under realistically tight evaluation budgets.

A few principles run through all four stages. Uncertainty is first-class, living in GEMD's distributional value types, in the open-source \texttt{Lolo} library's calibrated random-forest intervals~\cite{lolo}, and in the acquisition functions that drive sequential learning. Data are accepted as they come; the ``never turn away data'' posture deliberately lowers the barrier to adoption, and shared meaning is reconstructed afterward through template references rather than imposed from above. Models are judged by what they help discover, not by RMSE on a held-out set. And constraints, often treated elsewhere as filters applied after the fact, are treated here as part of the design problem itself.

The case studies in the Applications section show this stack working in practice. Sequential learning with well-calibrated uncertainty surfaced organic hole-transport semiconductors with mobilities beyond anything in the training set after only $\sim$165 DFT/MD evaluations~\cite{Antono2020}; it drove an autonomous closed-loop search for monodisperse colloidal palladium nanoparticles in tens of experiments instead of hundreds~\cite{Fong2021}; and it cut the number of measurements needed to find optima on magnetocaloric, superconducting, thermoelectric, and steel-fatigue benchmarks by factors of two to nine relative to random search~\cite{Ling2017}. Simulated discovery campaigns on top of those benchmarks make the converse point as well: better RMSE does not necessarily mean better discovery, and uncertainty-aware acquisition strategies are not optional~\cite{borg2023}.

Plenty of work remains. As autonomous laboratories~\cite{tabor2018automation,szymanski2023autonomous} get faster, the prediction--synthesis loop has to keep up, which puts pressure on every layer of the platform from data ingestion to acquisition latency. Multi-fidelity learning, where cheap simulation supports expensive experiment, can stretch sequential learning into regimes where data are still too scarce today. Generative and foundation models open up design spaces (especially for polymers) where simple enumeration is insufficient, but pairing them with calibrated uncertainty and hard constraints is still an open problem. And FAIR is far from solved: even with GEMD in hand, sustained community effort is needed to share template vocabularies and stand up the cross-institution infrastructure that the next decade of materials discovery will depend on~\cite{brinson2024fair,scheffler2022fair}.

Today the Citrine Platform regularly powers data-driven materials discovery across industries, having moved beyond one-off demonstrations into routine industrial practice. Getting there required solving a core set of recurring obstacles: scarce and inconsistent data, miscalibrated uncertainty, and constrained, high-dimensional design spaces. The Citrine Platform’s solution is not built on a single algorithm or standalone database, but on an integrated stack where data, modeling, and design-space layers continuously co-evolve. 
Continuing to develop those layers and the interactions between them, in close contact with both the materials science community and the emerging generation of autonomous experimentation systems, will drive the next decade of innovation in materials discovery.

\bibliography{bibliography}

\end{document}